\documentclass[aps,prd,onecolumn,superscriptaddress,nofootinbib]{revtex4-2}
\usepackage{graphicx,amsfonts,amsthm}
\usepackage{amsmath,amssymb,bbold,bm}
\usepackage{color}
\usepackage{booktabs}
\usepackage[utf8]{inputenc}

\newcommand{\doublet}[2]{ \left( \begin{array}{c}#1 \\ #2 \end{array}\right) }
\newcommand{\triplet}[3]{ \left( \begin{array}{c}#1 \\ #2 \\ #3 \end{array}\right) }

\newcommand{\lr}[1]{ \langle #1 \rangle}

\newcommand{\Z}{\mathbb{Z}}
\newcommand{\mmatrix}[4]{ \left(\! \begin{array}{ccc}#1 & #2 \\ #3 & #4 \end{array}\!\right) }

\providecommand{\mtrx}[1]{\begin{pmatrix} #1 \end{pmatrix}}

\newcommand{\id}{\mathbb{1}}
\newcommand{\lrpartial}{\,\partial^{\hspace{-7pt}\raise3pt\hbox{\small $\leftrightarrow$}}\!}
\newcommand{\scr}[1]{\mbox{\scriptsize #1}}
\newcommand{\VCKM}{V_{\mbox{\scriptsize CKM}}}
\newcommand{\hSM}{h_{\mbox{\scriptsize SM}}}
\newcommand{\mSM}{m_{\mbox{\scriptsize SM}}}

\def\lsim{\mathrel{\rlap{\lower4pt\hbox{$\sim$}}
    \raise1pt\hbox{$<$}}}         
\def\gsim{\mathrel{\rlap{\lower4pt\hbox{$\sim$}}
    \raise1pt\hbox{$>$}}}         

\begin{document}
\title{
Dark matter stabilized by a non-abelian group: lessons from the $\Sigma(36)$ 3HDM}

\author{Hong Deng}
\affiliation{School of Physics and Astronomy, Sun Yat-sen University,
519082 Zhuhai, China}
\author{Rafael Boto}\thanks{E-mail: rafael.boto@tecnico.ulisboa.pt}
\affiliation{CFTP, Instituto Superior T\'ecnico, Universidade de Lisboa,
	Lisbon, Portugal}
\author{Igor P. Ivanov}\thanks{E-mail: ivanov@mail.sysu.edu.cn}
\affiliation{School of Physics and Astronomy, Sun Yat-sen University,
519082 Zhuhai, China}
\author{Jo\~ao P. Silva}\thanks{E-mail: jpsilva@cftp.ist.utl.pt}
\affiliation{CFTP, Instituto Superior T\'ecnico, Universidade de Lisboa,
	Lisbon, Portugal}

\begin{abstract}
When building dark matter (DM) models, one often imposes conserved discrete symmetries 
to stabilize DM candidates. 
The simplest choice is $\Z_2$ but models with larger stabilizing groups have also been explored.
Can a conserved non-abelian group lead to a viable DM model?
Here, we address this question within the three-Higgs-doublet model based on the group $\Sigma(36)$,
in which DM stabilization by a non-abelian group is not only possible but inevitable.
We show that the tight connections between 
the Higgs, fermion, and DM sectors repeatedly drive the model into conflict 
with the LHC results and DM observations, with the most recent LZ results playing a decisive role.
We believe that the lessons learned from this study help chart the limits 
of what can be achieved within multi-Higgs-doublet DM models with large symmetry groups.
\end{abstract}

\maketitle

\section{Introduction}

\subsection{Stabilizing dark matter: symmetry groups vs. individual symmetries}

There exist several irrefutable pieces of evidence for dark matter (DM) filling the Universe \cite{Cirelli:2024ssz}. 
The mainstream view is that DM consists of one or several sorts of heavy, stable, electrically neutral particles of non-baryonic nature, 
which are absent in the Standard Model (SM) and, therefore, 
require some form of New Physics beyond the SM.
One popular idea is to assume that the scalar sector is not as minimal as in the SM but contains new scalar fields.
Equipped with a new global symmetry that remains unbroken in the vacuum, 
this scalar sector leads to one or more scalar DM candidates. 

A prototypical illustration of this scenario is the inert doublet model (IDM) \cite{Deshpande:1977rw,Ma:2006km,Barbieri:2006dq,LopezHonorez:2006gr}. 
This is a version of the two-Higgs-doublet model (2HDM) with an exact $\Z_2$ symmetry 
that acts trivially on the first Higgs doublet $\phi_1$ and the SM fields 
and flips the sign of the second Higgs doublet $\phi_2$.
If the vacuum expectation value (vev) of the second doublet is zero, $\lr{\phi_2} = 0$,
$\Z_2$ remains unbroken and stabilizes the lightest scalar from $\phi_2$ against decay.
As the model introduces very few free parameters and leads to numerous collider and cosmological predictions, 
it has become a popular playground, see for example 
\cite{Arhrib:2013ela,Ilnicka:2015jba,Belyaev:2016lok,Kalinowski:2018ylg,Belyaev:2018ext} and references therein.

Using $\Z_2$ as a DM stabilizing group is not the only option.\footnote{We mention in passing that 
there exist examples of scalar DM that are not based on any exact symmetry \cite{Kajiyama:2011gu}.}
One can easily construct multi-Higgs models in which DM candidates are protected by the unbroken $\Z_n$, $n > 2$, see 
\cite{Ma:2007gq,Batell:2010bp,Ivanov:2012hc,Belanger:2012vp,Belanger:2014bga,Yaguna:2019cvp,Belanger:2020hyh,Yaguna:2021vhb,Yaguna:2021rds,Belanger:2021lwd,Belanger:2022esk} and references therein.
Such models often contain additional processes that affect the DM evolution 
or feature several DM components with distinct interaction preferences.
One can even build a 3HDM model in which a pair of mass-degenerate DM candidates is stabilized 
by an exotic $CP$ symmetry of higher order rather than a Higgs family transformations \cite{Ivanov:2015mwl,Ivanov:2018srm}.
Another idea considered in literature is to replicate the IDM, that is, 
to consider a 3HDM with two independent $\Z_2$ symmetries flipping the signs of 
$\phi_2$ and $\phi_3$, respectively. 
The model possesses two DM candidates coming come from the two inert doublets 
with much freedom to adjust the properties of the two dark sectors
\cite{Belanger:2011ww,Aoki:2012ub,Bhattacharya:2016ysw,Hernandez-Sanchez:2020aop,Boto:2024tzp}. 

One may wonder what happens if scalar DM candidates are stabilized not by a single symmetry 
but by a non-abelian symmetry group. We stress that we talk about not a generic model whose lagrangian 
is invariant under a non-abelian global group $G$ (which can be broken later by the vevs),
but a special situation where a {\em remnant} non-abelian global symmetry group $G_v \subset G$ 
survives after the minimization of the scalar potential.
If this is the case, the DM candidates $h_i$ form a multiplet under $G_v$,
with additional channels
for relic density evolution, such as co-annihilation $h_i h_j \to \mbox{SM}$
or semi-annihilation $h_i h_j \to h_k+\mbox{SM}$. 
On the other hand, one expects that, due to many residual symmetries, 
these interactions are not fully arbitrary but correlated
with other observables of the model. 

This idea is not new. To our knowledge, the first---and the simplest---example based on the residual group $G_v = S_3$
was presented in \cite{Adulpravitchai:2011ei}. In that work, a simplified DM model was constructed,
which featured, in addition to the SM Higgs doublet, two electroweak (EW) singlet fields, the real $\eta$ 
and the complex $\chi$.
The interaction potential was designed in such a way that the model possesses $S_3$
and that the vev alignment preserves this group, making $\chi$ and $\chi^*$ the two mass degenerate DM candidates. 
The model contained enough freedom to introduce new channels
that affect the DM thermal evolution and to satisfy observational constraints.
Another attempt to build a DM model based on a non-abelian group, the quaternion group $Q_4$, 
was described in \cite{Lovrekovic:2012bz}.

In this paper, we are interested in a different twist of the non-abelian DM scenario.
In the above papers, as well as in many other DM models, one postulates a dark sector which, 
by construction, does not participate in the EW interactions and couples to the visible sector
either through portal-like interactions or flavor physics connections. 
Since the visible and the dark sectors are governed by different lagrangians, 
it is no wonder that this scenario offers enough freedom
to adjust the DM sector separately from the SM fields.

However, we want to explore the same scenario in the multi-Higgs-doublet models.
Here, the scalar DM protected by a non-abelian group appears from the same self-interaction
potential as the SM-like Higgs boson itself. As a result, the connections and constraints become much tighter,
and it is not clear a priori whether such a DM scenario is tenable. 
What we want is not to simply construct yet another DM model but to chart the limits 
of what multi-Higgs-doublet models can in principle accommodate.
Can they incorporate such a scenario? Does it require much fine-tuning or any additional assumptions? 
Does it run into immediate conflicts with collider and cosmological observations?
Does it lead to specific predictions?

When addressing these questions, we make as few assumptions as possible and check where the analysis leads us.
It may seem that a residual non-abelian symmetry group of a multi-Higgs-doublet model 
must hinge on several assumptions. 
This is not so. In fact, we start with only one assumption: we choose a multi-Higgs-doublet potential 
which is invariant under a large finite group $G$ of global symmetries.
Experience shows that if $G$ is sufficiently large---and it of course 
depends on what ``sufficiently'' means---non-abelian DM follows automatically.
Namely, it turns out that, for a sufficiently large $G$, {\em any} minimum of the scalar potential 
keeps unbroken some non-abelian subgroup $G_v \subset G$.
Thus, DM stabilization by a non-abelian residual symmetry group
is not only possible but inevitable in such models.
It was this initial observation that motivated the present study.

In this work, we explore this phenomenon and its DM implications with the example of the three-Higgs-doublet model (3HDM)
whose scalar sector is invariant under the global symmetry group $G=\Sigma(36)$.
This option first appeared in the systematic classification of the finite realizable symmetry groups
of the 3HDMs \cite{Ivanov:2012ry,Ivanov:2012fp}; 
the fact that all its global minima preserve a non-abelian subgroup $G_v = S_3 \subset \Sigma(36)$ 
was mentioned in \cite{Ivanov:2014doa}.
Recently, it was shown in \cite{Yang:2024bys} that, with a softly broken $\Sigma(36)$,
this model can accommodate cosmological phase transition scenarios, in which
expanding bubbles of true vacuum are separated from the false vacuum background by charge-breaking bubble walls.
The $S_3$-stabilized DM evolution, which we study here, is another interesting feature of this model.

\subsection{The objectives and the structure of the paper}

The main objective of this paper is to explore the characteristic features of the DM 
in multi-Higgs-doublet models highly constrained by large global symmetry groups.
We will see that it is not an easy task to arrange for stable DM candidates that could account for at least a part 
of the total relic density without running into conflict with direct detection and collider constraints.
We believe that the $\Sigma(36)$-based 3HDM offers us lessons about this general phenomenon.

The sequence of the steps we will take along these lines is the following.
In the next section, we will review the scalar sector of the $\Sigma(36)$ 3HDM,
list its minima and the physical scalars.
We will also highlight the features that arise due to stabilization of DM candidates
by a non-abelian group, as contrasted with stabilization by a single symmetry.
Then, in Section~\ref{section-realistic} we attempt to complete the model
with a suitable Yukawa sector.
We first show in Section~\ref{section-Yukawa-exact} that $\Sigma(36)$ can be, 
in principle, extended to the quark sector but leads to unphysical quark properties.
Even this situation would be acceptable as a toy model, had we been able to keep the DM candidates stable.
But it turns out that all possible $\Sigma(36)$-invariant Yukawa sectors
unavoidable lead to scalar decays to quark pairs, precluding any stable DM scenario. 
This no-go result is the first lesson we learn.

We then relax our assumptions and, in Section~\ref{section-Yukawa-S3}, we construct a Yukawa sector that
is invariant not under the full $\Sigma(36)$ but under its subgroup $G_v = S_3$,
the residual symmetry group at a chosen minimum.
This can be easily done in a fully realistic way.
Moreover, since $\Sigma(36)$ is not the symmetry of the full lagrangian,
we also take the liberty of introducing soft breaking terms respecting the same $S_3$,
which adds only one additional parameter.
This is done in a consistent way, so that the minimum we initially selected
becomes the unique global minimum of the resulting potential.

We incorporate the model in micrOMEGAs \cite{Alguero:2023zol} and present, 
in Section~\ref{section-DM}, the results of our study.
First, we focus on the model without any soft breaking and
explore how the relic density and direct detection signals depend on the free parameters.
We compare our predictions with the Planck observations \cite{Planck:2018vyg}
and the upper limits from the direct detection experiments Xenon1T \cite{XENON:2018voc} and LZ \cite{LZ:2022lsv},
including the very recent LZ results \cite{LZCollaboration:2024lux}. 
We find that, due to a strong correlation among the signals, the $\Sigma(36)$-symmetric scalar sector 
is unable to account even for a part of DM density. 
This strong conclusion is the second lesson emerging from our study.

We then proceed to the softly broken $\Sigma(36)$ 3HDM, 
complemented with the $S_3$ symmetric Yukawa sector,
and discuss in Section~\ref{subsection-DM-soft} the DM relic density and 
direct detection signals.
As the tight correlation between the scalar masses and couplings is now relaxed,
we find it possible to match the DM abundance or to remain in the sub-dominant DM scenario
without violating observational constraints.
However, even with these relaxed assumptions, we see a serious clash between theoretical constraints 
and the LZ-2024 direct detection results, which is the third lesson of our study.
We place a lower bound on the magnitude of the soft breaking terms
that allow us to reach minimally viable models, and comment on the origin of these persistent conflict 
between different experimental constraints.

\section{The scalar sector of the $\Sigma(36)$ 3HDM}\label{section:general-remarks}

\subsection{Structural rigidity and its consequences}\label{subsection:Delta27}

Structural rigidity is the central theme of the symmetry-based multi-Higgs model building activity.
Here, ``rigidity'' means that, for a sufficiently large finite group, the phenomenological properties of the model
follow robust patterns insensitive to the exact numerical values of the parameters.
The key goal is then to identify which pattern seems to (approximately) fit the observables
and which multi-Higgs models lend support to such a pattern.

We find it instructive to remind the reader of the situation in which the particle physics found itself
at the end of the 1970s. With the third generation of fermions just discovered, 
and their masses and mixing parameters measured, 
the hierarchical structure of the fermion masses and the Cabibbo-Kobayashi-Maskawa (CKM) 
matrix $\VCKM$ called for explanations. The problem was optimistically attacked by constructing 
$N$-Higgs-doublet models with several generations of Higgs doublets. The scalar doublets were assumed to couple to fermions and among themselves
in such a way that the lagrangian stayed invariant under a group $G$ of global transformations acting both on the scalar and fermions
fields. In this early activity, several options for the group $G$ were studied, such as the groups of sign flips, 
rephasing transformations, permutations, or their combinations; for a brief historical overview see \cite{Ivanov:2017dad}.
For example, the early suggestion \cite{Segre:1978ji} 
was based on the group of $\Delta(27) \subset SU(3)$ generated by the following order-3
rephasing and permutation transformations: 
\begin{equation}
	a = \mtrx{1&0&0\\ 0&\omega&0\\ 0&0&\omega^2}\,, \quad
	b = \mtrx{0&1&0\\ 0&0&1\\ 1&0&0}\,,\quad\mbox{where}\quad \omega = e^{2\pi i /3}, \ \omega^3 = 1\,.
	\label{Delta27-generators}
\end{equation}
The idea of Ref.~\cite{Segre:1978ji} was to ``naturally'' explain the mass and mixing hierarchies by assuming that 
the three Higgs doublets acquire hierarchical vevs $|v_1| \ll |v_2| \ll |v_3|$.
However, as it was pointed out already in \cite{Segre:1978ji}, the $\Delta(27)$ invariant scalar potential,
with its limited structures and very few free parameters, was unable to yield such a vev alignment. 

The failure of the $\Delta(27)$-invariant 3HDM to reproduce the desired vevs was later understood as a particular example of a general phenomenon.
Multi-Higgs potentials based on large finite symmetry groups possess a rigid structure of their minima \cite{Degee:2012sk,Ivanov:2014doa}
and even saddle points \cite{Yang:2024bys}. Their vevs scale as simple ratios 
such as $1:1:1$ and remain stable against smooth variation of the coefficients.
Another manifestation of a rigid vev structure is the so-called geometric $CP$ violation,
the calculable relative phase between vevs,
which was first noticed in \cite{Branco:1983tn} for the same $\Delta(27)$ 3HDM and 
later used in other multi-Higgs models \cite{deMedeirosVarzielas:2011zw,deMedeirosVarzielas:2012rxf,Ivanov:2013nla}.

Structural rigidity has far-reaching consequences for the fermion sector of the model.
As it was first pointed out in \cite{Leurer:1992wg}, 
structural rigidity may render the quark sector unphysical at any minimum of a symmetry-constrained potential. 
A no-go theorem was formulated in \cite{Leurer:1992wg} and later refined in \cite{GonzalezFelipe:2014mcf}, 
which states that the only way to obtain a non-block-diagonal CKM mixing matrix and, simultaneously, non-degenerate
and non-zero quark masses is to make sure that vev alignment breaks the group $G$ completely, except possibly for a symmetry
belonging to the baryon number or a symmetry located purely within the inert sector (i.e. for the doublets which do not couple to quarks).
But in potentials with a large symmetry group, there are always residual symmetries left at each of the possible minima.
One popular way out is to add soft breaking terms, which can in many cases provide enough freedom to adjust vevs
at least within some limits.
A nice illustration of the above situation was given for the $A_4$-invariant 3HDM:
while the model with an exact $A_4$ symmetry always leads to pathological fermion sectors
\cite{GonzalezFelipe:2013xok}, adding soft breaking terms was enough
to remove the unwanted residual symmetries and render the model compatible with experiment \cite{Bree:2023ojl}.
For the $\Delta(27)$ symmetric 3HDM, numerous attempts were undertaken 
to incorporate it into realistic models \cite{deMedeirosVarzielas:2011zw, Varzielas:2012nn, Varzielas:2013eta, Kalinowski:2021lvw},
but all of them had to resort to additional non-SM fields, going beyond the pure multi-Higgs-doublet models,
or even had to accept the massless fermions as a viable option as in \cite{Kalinowski:2021lvw}.

Stabilization of scalar DM candidates is another manifestation of structural rigidity,
as certain vevs can remain exactly zero in broad regions of the parameter space.
Thus, rigidity can be either a welcome feature, as in the case of DM candidates, 
or a nuisance factor, as for the fermion sector.
Whether one can keep its positive features and avoid its downsides remains to be analyzed in specific models.

\subsection{The $\Sigma(36)$ 3HDM potential}\label{subsection:potential}

Let us now describe and analyze the $\Sigma(36)$ 3HDM scalar sector. 
First, a technical remark is in order. The group $\Sigma(36)$ derived in the 3HDM symmetry classification 
\cite{Ivanov:2011ae,Ivanov:2012ry,Ivanov:2012fp} was, by constructed, considered as a subgroup of $PSU(3) \simeq SU(3)/Z(SU(3))$,
where  $Z(SU(3)) \simeq \Z_3 = 
\{\id_3, \, \omega\cdot \id_3, \, \omega^2\cdot \id_3\}$ is the center of $SU(3)$, see more discussion in Appendix~\ref{appendix-technical}.
However, since it is convenient to work with unitary, not projective representations, 
we will work inside $SU(3)$. In this group space, the finite symmetry group we deal with is denoted as $\Sigma(36\varphi)$, or $\Sigma(36\times 3)$, 
see e.g. \cite{Grimus:2010ak,Hagedorn:2013nra}, and is of order 108. 
Although we will occasionally write $\Sigma(36)$ as in ``$\Sigma(36)$-based 3HDM,'' we will actually deal with $\Sigma(36\varphi)$.

The group $\Sigma(36\varphi)$, which can also be represented as $\Delta(27)\rtimes \Z_4$, is generated by the same $a$ and $b$ as 
in Eq.~\eqref{Delta27-generators} and a new transformation $d$ of order four:
\begin{equation}
	d = \frac{i}{\sqrt{3}} \left(\begin{array}{ccc} 1 & 1 & 1 \\ 1 & \omega^2 & \omega \\ 1 & \omega & \omega^2 \end{array}\right)\,,
	\quad
	\mbox{so that}\  d^2 = -\mtrx{1&0&0\\ 0&0&1\\ 0&1&0}\,.
	\label{Sigma36-generators}
\end{equation}
The 3HDM scalar potential invariant under this group is
\begin{eqnarray}
	V_0 & = &  - m^2 \left(\phi_1^\dagger \phi_1+ \phi_2^\dagger \phi_2+\phi_3^\dagger \phi_3\right)
	+ \lambda_1 \left(\phi_1^\dagger \phi_1+ \phi_2^\dagger \phi_2+\phi_3^\dagger \phi_3\right)^2 \nonumber\\
	&&
	- \lambda_2 \left[|\phi_1^\dagger \phi_2|^2 + |\phi_2^\dagger \phi_3|^2 + |\phi_3^\dagger \phi_1|^2
	- (\phi_1^\dagger \phi_1)(\phi_2^\dagger \phi_2) - (\phi_2^\dagger \phi_2)(\phi_3^\dagger \phi_3)
	- (\phi_3^\dagger \phi_3)(\phi_1^\dagger \phi_1)\right] \nonumber\\
	&&
	+ \lambda_3 \left(
	|\phi_1^\dagger \phi_2 - \phi_2^\dagger \phi_3|^2 +
	|\phi_2^\dagger \phi_3 - \phi_3^\dagger \phi_1|^2 +
	|\phi_3^\dagger \phi_1 - \phi_1^\dagger \phi_2	|^2\right)\, .
	\label{Vexact}
\end{eqnarray}
Here, the first two lines are invariant under the entire $SU(3)$ group of family rotations of the three Higgs doublets,
while the $\lambda_3$ term selects the finite group $\Sigma(36\varphi)$.
This potential is also $CP$ invariant \cite{Ivanov:2011ae,Ivanov:2014doa};
in fact, the $\Z_4$ symmetry group within the 3HDM scalar sector automatically forbids 
any form of $CP$ violation in the scalar sector, be it explicit or spontaneous.

The potential in Eq.~\eqref{Vexact} contains very few free parameters.
The coefficients $m^2$ and $\lambda_1$ fix the overall scales 
of the vev $v$ and the SM-like Higgs mass,
while the locations of the global minima depend on $\lambda_2$ and $\lambda_3$.
The conditions for the potential to be bounded from below (BFB) were derived in \cite{Yang:2024bys}:
\begin{equation}
	\lambda_1 > 0\,, \quad \lambda_1 + \lambda_3 > 0\,,\quad
	\lambda_1 + \frac{1}{4}\lambda_2 > 0\,,\quad \lambda_1 + \frac{1}{4}(\lambda_2 + \lambda_3) > 0\,.
	\label{BFB}
\end{equation}
All these inequalities are strict, meaning we must impose the BFB conditions in the strong sense \cite{Maniatis:2006fs}.

\subsection{The global minima and residual symmetries}

Depending on the values of the quartic coefficients, the global minimum of the $\Sigma(36)$ 3HDM can be either neutral or charge-breaking. 
The latter option must be avoided at zero temperature, although existence of a charge-breaking phase
at intermediate temperatures is an interesting and insufficiently explored opportunity
for the hot early Universe evolution \cite{Aoki:2023lbz,Yang:2024bys}. 
The regions on the $(\lambda_2,\lambda_3)$ plane that guarantee the neutral minimum of the tree-level potential 
were established in \cite{Yang:2024bys} and are listed below.

The vev alignment corresponding to a neutral global minimum can be written as
\begin{equation}
	\lr{\phi_1} = \frac{1}{\sqrt{2}}\doublet{0}{v_1}\,,\quad  
	\lr{\phi_2} = \frac{1}{\sqrt{2}}\doublet{0}{v_2}\,,\quad 
	\lr{\phi_3} = \frac{1}{\sqrt{2}}\doublet{0}{v_3}\,,
	\label{neutral-vevs}
\end{equation}
with $v_i$ being, in general, complex. To simplify the notation, we will take the overall scale out, 
implicitly assuming that $|v_1|^2 + |v_2|^2 + |v_3|^2 = v^2$, where $v = 246$~GeV, 
and indicate the ratios between the individual vevs. 
For example, if a minimum corresponds to $v_1 = v_2 = v_3$, we write 
\begin{equation}
	(\lr{\phi_1^0},\, \lr{\phi_2^0},\, \lr{\phi_3^0}) = 
	\left(\frac{v_C}{\sqrt{2}},\, \frac{v_C}{\sqrt{2}},\, \frac{v_C}{\sqrt{2}}\right) = \frac{v_C}{\sqrt{2}}\left(1, 1, 1\right)\,,
	\quad
	\mbox{with}\quad v_C = \frac{v}{\sqrt{3}}\,,
\end{equation}
and label this vev alignment as $(1,1,1)$.

Depending on the free parameters, $\Sigma(36)$ 3HDM allows for two scenarios for neutral minima \cite{Ivanov:2014doa,Yang:2024bys}.
\begin{itemize}
	\item 
	$BC$-scenario: if $\lambda_2 > 0$ and $\lambda_3 > 0$, then the global minima correspond to the following vev alignments: 
	\begin{eqnarray}
		&&	\mbox{alignment $B$:}\quad B_1 = (1,\,0,\,0)\,, \quad B_2 = (0,\,1,\,0),\, \quad B_3 = (0,\,0,\,1)\label{points-B}\\
		&&\mbox{alignment $C$:}\quad C_1 = (1,\,1,\,1)\,,\quad C_2 = (1,\,\omega,\,\omega^2)\,,\quad C_3 = (1,\,\omega^2,\,\omega)\label{points-C}\,.
	\end{eqnarray}
	Other configurations such as $(\omega,\,\omega^2,\, 1)$ can be 
	reduced to those already listed by an overall phase rotation of the three doublets.
	For example, $(\omega,\,\omega^2,\, 1) = \omega (1,\,\omega,\,\omega^2)$ corresponds to the alignment $C_2$.
	\item 
	$AA'$-scenario: if $\lambda_3 < 0$, while $\lambda_2$ satisfies $\lambda_2 - 3 \lambda_3 > 0$,
then the global minima correspond to
	\begin{eqnarray}
		&&\mbox{alignment $A$:}\quad A_1 = (\omega,\,1,\,1)\,, \quad A_2 = (1,\,\omega,\,1),\, \quad A_3 = (1,\,1,\,\omega)\label{points-A}\\
		&& \mbox{alignment $A'$:}\quad A'_1 = (\omega^2,\,1,\,1)\,, \quad A'_2 = (1,\,\omega^2,\,1),\, \quad A'_3 = (1,\,1,\,\omega^2)\label{points-Ap}
	\end{eqnarray}
\end{itemize}
In either case, we observe a similar picture: there are six degenerate global minima, which are all linked by the transformations from the group $\Sigma(36)$
broken by the vev alignment. For example, the broken transformation $d$ links $B_i$ and $C_i$.
Also, no higher-lying local minima exists for any of these cases \cite{Yang:2024bys}.

For any choice of the global minimum, the symmetry group $G = \Sigma(36)$ is spontaneously 
broken to the residual subgroup $G_v = S_3$. 
It is well known that any $S_3$ can be generated by two generators $g_2$ and $g_3$ satisfying $g_2^2 = g_3^3 = e$, $g_2^{-1} g_3 g_2 = g_3^{-1}$.  
The specific transformations $g_2$ and $g_3$ from $\Sigma(36)$ that remain unbroken and generate this residual $S_3$ differ for each minimum.
Here are several examples of the residual symmetry generators:
\begin{eqnarray}
	&&\mbox{alignment}\ B_1 = (1,\,0,\,0):\quad g_2 = d^2\,, \quad g_3 = a\,.\label{residual-B1}\\
	&&\mbox{alignment}\ C_1 = (1,\,1,\,1):\quad  g_2 = d^2\,, \quad g_3 = b\,.\label{residual-C1}\\
	&&\mbox{alignment}\ C_2 = (1,\,\omega,\,\omega^2):\quad g_2 = d^2a\,, \quad g_3 = b\,.\label{residual-C2}\\
	&&\mbox{alignment}\ A_1 = (\omega,\,1,\,1):\quad g_2 = d^2\,, \quad g_3 = ba^2\,.\label{residual-A1}
\end{eqnarray}
Besides, each vev alignment conserves $CP$, either the canonical $CP_0: \phi_a \mapsto \phi_a^*$
or a generalized $CP$ symmetry which combines $CP_0$ with a transformation from $\Sigma(36)$.
For example, the alignment $A_1$ is invariant under $CP_0$ followed by $d$.

\subsection{The physical scalars and DM candidates}

Since any choice of the global minimum leads to residual symmetries, we expect scalar DM candidates stabilized by these symmetries.
The fact that the residual symmetry group is $S_3$ hints at a two-dimensional space 
of mass-degenerate DM candidates. This is the feature which we want to explore with our choice of the symmetry group $\Sigma(36)$.

In this subsection, we give the spectrum of physical scalars and indicate their conserved quantum numbers which protect 
them against decay. We remind the reader that, at this stage, we are working with the bosonic degrees of freedom only.

Let us begin with the $BC$-scenario and choose the vev alignment $B_1$ as a representative case. 
We expand the Higgs doublets around the minimum as
\begin{equation}
	\phi_1 = \frac{1}{\sqrt{2}}\doublet{\sqrt{2}h_1^+}{v + h_1 + i a_1}\,,\quad  
	\phi_2 = \frac{1}{\sqrt{2}}\doublet{\sqrt{2}h_2^+}{h_2 + i a_2}\,,\quad 
	\phi_3 = \frac{1}{\sqrt{2}}\doublet{\sqrt{2}h_3^+}{h_3 + i a_3}\,.
	\label{field-expansion}
\end{equation}
The fields $h_1^+$ and $a_1$ are the would-be Goldstone bosons which disappear in the unitary gauge.
The remaining fields are physical scalars with the following masses:
\begin{eqnarray}
	h_{SM} = h_1: &\qquad & m_{SM}^2 = 2\lambda_1 v^2 = 2 m^2\,,\label{B1-spectrum-SM}\\
	H_2^+ = h_2^+\,, \ H_3^+ = h_3^+: &\qquad & m_{H^+}^2 = \frac{\lambda_2 v^2}{2}\quad \mbox{(double degenerate)}\,,\label{B1-spectrum-charged}\\
	h = \frac{1}{\sqrt{2}}(h_2 + h_3)\,, \quad a = \frac{1}{\sqrt{2}}(a_2 - a_3): &\qquad &
	m_{h}^2 = \frac{\lambda_3 v^2}{2}\quad \mbox{(double degenerate)}\,,\label{B1-spectrum}\\
	H = \frac{1}{\sqrt{2}}(h_2 - h_3)\,, \quad A = \frac{1}{\sqrt{2}}(a_2 + a_3): &\qquad &
	m_{H}^2 = \frac{3 \lambda_3 v^2}{2}\quad \mbox{(double degenerate)}\,.\label{B1-spectrum-HA}
\end{eqnarray}
Assuming $\lambda_2 > \lambda_3$, the neutral fields $h$ and $a$ indicated in the third line
are the DM candidates.
Since they are mass degenerate, one can alternatively choose any linear combination of 
$h$ and $a$, which can also be considered as the DM candidate.
Indeed, the residual symmetry group $S_3$ given in Eq.~\eqref{residual-B1}
acts on $h$ and $a$ by $2\pi/3$ rotations and reflections in the $(h,a)$ plane.
Thus, the pair $(h,a)$ forms a real 2D irreducible representation of $S_3$, and it is this 2D space of spaces
which is stabilized by the residual group.

For the $AA'$-scenario, we choose the alignment $A_3 = (1,1,\omega)$ and parametrize the doublets as
\begin{equation}
	\phi_1 = \frac{1}{\sqrt{2}}\doublet{\sqrt{2}h_1^+}{v_A + h_1 + i a_1}\,,\quad  
	\phi_2 = \frac{1}{\sqrt{2}}\doublet{\sqrt{2}h_2^+}{v_A + h_2 + i a_2}\,,\quad 
	\phi_3 = \frac{\omega}{\sqrt{2}}\doublet{\sqrt{2}h_3^+}{v_A + h_3 + i a_3}\,,
	\label{field-expansion-2}
\end{equation}
where $v_A = v/\sqrt{3}$. Note that we took the phase factor $\omega$ out of the third doublet.
The mass of the SM-like Higgs is again
\begin{equation}
	\hSM = \frac{1}{\sqrt{3}}(h_1 + h_2 + h_3): \quad \mSM^2 = 2(\lambda_1 + \lambda_3) v^2 = 2m^2\,. 
\end{equation}
The charged Goldstone is $G^+ = (h_1^+ + h_2^+ + h_3^+)/\sqrt{3}$, while the two charged Higgses orthogonal to it
are again mass degenerate:
\begin{equation}
	H_2^+, \, H_3^+ \ \perp \ G^+: \quad m_{H^+}^2 = \frac{v^2}{2}(\lambda_2 - 3\lambda_3)\,. 
\end{equation}
To find the remaining four neutral mass eigenstates, we define
\begin{equation}
	h_- = \frac{1}{\sqrt{2}}(h_1 - h_2)\,, \quad 
	h_+ = \frac{1}{\sqrt{6}}(h_1 + h_2 - 2 h_3)\,, \quad
	a_- = \frac{1}{\sqrt{2}}(a_1 - a_2)\,, \quad 
	a_+ = \frac{1}{\sqrt{6}}(a_1 + a_2 - 2 a_3)\,,
\end{equation}
and observe that the neutral mass matrix splits into two identical $2\times 2$ blocks ${\cal M}_2$ acting within the spaces
$(h_-,a_-)$ and $(h_+, a_+)$:
\begin{equation}
	{\cal M}_2 = -\frac{\lambda_3 v^2}{4} \mmatrix{5}{\sqrt{3}}{\sqrt{3}}{3}\,.
\end{equation}
Inside $(h_-,a_-)$, the eigenvectors and eigenvalues are
\begin{eqnarray}
	h = -\frac{1}{2}h_- + \frac{\sqrt{3}}{2}a_-\,: &\quad& m_h^2 = -\frac{\lambda_3 v^2}{2}\nonumber\\
	H = \frac{\sqrt{3}}{2}h_- + \frac{1}{2}a_-\,: &\quad& m_H^2 = -\frac{3\lambda_3 v^2}{2}\,.
\end{eqnarray}
Note that $A,A'$ can be the minima only for $\lambda_3 < 0$. 
We observe the same mass hierarchy $m_H^2 = 3 m_h^2$ as in Eqs.~\eqref{B1-spectrum},~\eqref{B1-spectrum-HA}.
The construction of the physical states in the space $(h_+, a_+)$ proceeds in the same way,
yielding $a$ and $A$. Thus, we again have two mass degenerate DM candidates, $h$ and $a$.

\subsection{DM stabilization by a group vs. by a single symmetry}

In the above discussion, we state that the DM candidates are stabilized by the residual symmetry group $S_3$
rather than by individual residual symmetries.
This language may be somewhat unconventional but it is fully justified. 
Suppose that, trying to stick to the conventional language,
we would say that $a$ is stabilized by the residual $\Z_2$ symmetry $-g_2$, which flips the sign of $a$
but keeps $h$ unchanged.
Then we would not be able to identify the symmetry which stabilizes $h$ alone.
Alternatively, we could switch to the complex fields
\begin{equation}
	\chi = \frac{1}{\sqrt{2}}(h + i a)\,, \quad \chi^* = \frac{1}{\sqrt{2}}(h - i a)\label{psipsi}
\end{equation}
and say that they are stabilized by the residual $\Z_3$ symmetry, under which they transform with the charges 
$+1$ and $-1$. Although such a representation may be useful to identify the semi-annihilation channel
$\psi \psi \to \psi^*+\mbox{SM}$, it still does not capture the full picture.
The full information on DM stabilization is provided by indicating that it is the group $S_3$
that protects the two real component scalar DM candidates.


Stabilization by a non-abelian group leads to another important feature, 
which prevents an immediate clash with experiment that other models 
with two mass-degenerate scalar DM candidates may possess.
Consider again the IDM, the 2HDM with the residual symmetry group $\Z_2$.
In this model, $H$ is the lightest DM candidate stabilized by $\Z_2$, 
and $A$ is a heavier state, which is also $\Z_2$ odd.
Being members of the same doublet and possessing the opposite $CP$ parities, 
the two scalars can interact through the $ZHA$ interaction:
\begin{equation}
	{\cal L} \supset i \frac{g}{2 c_W}\, Z^\mu\left(H\partial_\mu A - A\partial_\mu H\right)\,.\label{ZHA}
\end{equation}
Then, in the mass-degenerate limit of $m_A = m_H$, we find that the DM particles
can elastically scatter off a nucleus via the transitions $H \to A$ and $A\to H$
mediated by the $Z$-exchange.
This scattering would lead to an unacceptable large direct detection (DD) signal
and is definitely ruled out by the negative results of DD searches.

One may wonder if the two degenerate DM candidates $h$ and $a$, which our model contains,  
can also couple to the $Z$ bosons in a similar way.
Fortunately, they do not; we observe only the $ZhA$ and $ZHa$ vertices but not $Zha$. 
The reason is again group-theoretical. Since $(h, a)$ form a doublet under $S_3$,
their hypothetical interaction terms with $Z$ 
must be of the type $\bm{2} \otimes \bm{2} = \bm{1} \oplus \bm{1}' \oplus \bm{2}$.
Since the conserved $S_3$ does not act on the $Z$, we must select the trivial singlet, 
which involves only the diagonal combinations.
But the $Zhh$ and $Zaa$ couplings are impossible by construction,
and the non-symmetric $Zha$ does not enter the trivial $S_3$ singlet.
We conclude that the residual group $S_3$ protects the model against the $Zha$ interactions.

Finally, stabilization of DM by $S_3$ differs from the case of $\Z_2\times\Z_2$ in that it allows 
for semi-annihilation channels such as $hh \to a+\mbox{SM}$.
This is due to the fact that the $S_3$ decomposition 
of $\bm{2} \otimes \bm{2}$, which corresponds to the $S_3$ quantum numbers of the initial state, 
contains a $\bm{2}$, which matches the final state quantum numbers.


\section{Building realistic $\Sigma(36)$-based dark matter models}\label{section-realistic}

\subsection{A toy model: exact $\Sigma(36)$ in the Yukawa sector}\label{section-Yukawa-exact}

In order to track the freeze-out evolution of DM density and explore the interplay of two DM candidates, 
one must complement the scalar interactions with a minimally realistic Yukawa sector.
In this section, we describe our attempt to extend the full symmetry group $\Sigma(36)$ to the quark Yukawa.
The no-go theorem proved \cite{Leurer:1992wg,GonzalezFelipe:2014mcf} makes it clear that,
even if such a sector can be constructed, the resulting quark properties 
cannot fully match the experimentally measured values and will unavoidably display pathologies 
such as massless quarks and insufficient mixing parameters.
Nevertheless, since the main goal of our study is to check the evolution of scalar DM stabilized 
by a non-abelian residual group, we find it appropriate to embrace this situation, considering it as a toy model,
provided the scalar DM candidates remain stable.

To set up notation, we write the quark Yukawa sector of the 3HDM as
\begin{equation}
	-{\cal L}_Y = \overline{Q}^0_{Li} \Gamma_{a,ij} \phi_a d_{Rj}^0 +
	\overline{Q}^0_{Li} \Delta_{a,ij} \tilde\phi_a u_{Rj}^0 + h.c.\label{Yukawa-general}
\end{equation}
Here, the indices $i,j = 1,2,3$ refer to the quark generations, while $a = 1,2,3$ label the Higgs doublets. 
The superscript $0$ for the quark fields indicates that these are the starting quark fields;
when we pass to the physical quarks by diagonalizing their mass matrices, we will remove this superscript.

To construct the $\Sigma(36)$-invariant quark Yukawa sector, we need to assign the quark fields $Q_L$,
$d_R$, and $u_R$ to specific group representations, follow the rules for decomposition of their products,
and extract the trivial singlet.
To do this, we first briefly review the irreducible representations (irreps) of $\Sigma(36\varphi)$, 
reproducing some of the results of \cite{Grimus:2010ak,Hagedorn:2013nra}.

The finite group $\Sigma(36\varphi)$ has four 1D irreps labeled $\bm{1}^{(p)}$, $p = 0, 1, 2, 3$,
which trivially represent the generators $a$ and $b$, $\rho_{1^{(p)}}(a) = \rho_{1^{(p)}}(b) = 1$,
and differ only by the representing value of the generator $d$: $\rho_{1^{(p)}}(d) = i^p$.
For shorthand notation, we will denote the trivial representation $\bm{1}^{(0)}$ as $\bm{1}$.
Next, it has four complex 3D irreps $\bm{3}^{(p)}$, whose $\rho_3(a)$ and $\rho_3(b)$ are given by Eq.~\eqref{Delta27-generators}
and whose $\rho_3(d)$ is as in Eq.~\eqref{Sigma36-generators} with the same extra factor $i^{p}$.
These are complemented by the four conjugate triplets $(\bm{3}^{(p)})^*$. 
Finally, the group possesses two real 4D irreps, $\bm{4}$ and $\bm{4}'$, which we will not encounter in our construction.
The squares of the irrep dimensions sum up to the group order as $\sum_i d_i^2 = 4\times 1 + 8\times 3^2 + 2\times 4^2 = 108$.

In Appendix~\ref{appendix-group}, we collect the irrep decomposition rules for the products of a triplet with
another triplet or an antitriplet. These rules follow a uniform pattern, which allows us, without losing generality, 
to assign the Higgs doublets to $\bm{3}^{(0)}$, which we shorten as $\bm{3}$.
Then, the two most relevant decomposition rules are
\begin{equation}
	\bm{3} \otimes \bm{3}^* = \bm{1} \oplus \bm{4} \oplus \bm{4}'\,, \qquad
	\bm{3} \otimes \bm{3} = \bm{3}^* \oplus (\bm{3}^{(1)})^* \oplus (\bm{3}^{(3)})^*\,.\label{irrep-products}
\end{equation}
It is instructive to compare this situation with the irreps of $\Delta(27)$ and their product decomposition,
which can be found, for example, in \cite{deMedeirosVarzielas:2015amz}.
The group $\Delta(27)$ has nine distinct 1D irreps $\bm{1}_{ij}$,
which are labeled by the powers of $\omega$ in their $\rho_1(a)$ and $\rho_1(b)$,
and two complex 3D irreps, $\bm{3}$ and $\bm{3}^*$.
As a result, $\bm{3} \otimes \bm{3}^*$ decomposes into the sum of all nine distinct singlets.
This feature makes the group convenient for flavor model building. Indeed, assuming that $\phi_a$ and $Q_{Li}$
transform as $\bm{3}$, one can assign each individual $d_{Rj}$ to a different singlet, 
which introduces flexibility in building the Yukawa sector \cite{Kalinowski:2021lvw}.
In the case of $\Sigma(36\varphi)$, this freedom is strongly reduced, because 
the product $\bm{3} \otimes \bm{3}^*$ involves now only one singlet. 

In Appendix~\ref{appendix-group}, we list all classes of irrep assignments for the quark fields.
Although the Yukawa matrices $\Gamma_a$ and $\Delta_a$ depend on cases,
the consequences are similar: in each sector, we obtain massless or mass degenerate quarks and insufficient mixing.
For example, in the case where $Q_L$ are trivial singlets, $d_R$ transform as $\bm{3}^*$, 
and $u_R$ transform as $\bm{3}$, 
the three Yukawa matrices $\Gamma_a$ take the following form:
\begin{equation}
	\Gamma_1 = \mtrx{g_1&\cdot&\cdot \\ g_2&\cdot&\cdot \\ g_3&\cdot&\cdot}\,, \quad 
	\Gamma_2 = \mtrx{\cdot&g_1&\cdot \\ \cdot&g_2&\cdot \\ \cdot&g_3&\cdot}\,, \quad 
	\Gamma_3 = \mtrx{\cdot&\cdot&g_1 \\ \cdot&\cdot&g_2 \\ \cdot&\cdot&g_3}\,,\label{case1333-Gamma}
\end{equation}
where dots correspond to the zero entries. The three independent coefficients $g_i$ are in general complex.
In the up-quark sector, the matrices $\Delta_a$ have the same structure,
bearing their own parameters $d_i$. The quark mass matrices become rank-1 matrices: 
$(M_d)_{ij} = g_i v_j/\sqrt{2}$, $(M_u)_{ij} = d_i v^*_j/\sqrt{2}$.
Diagonalizing these matrices, we find two generations of massless quarks in the down and up-quark sectors
and one generation of massive ones, with $m_b^2 = v^2 |\vec g|^2/2$ and $m_t^2 = v^2 |\vec d|^2/2$.
These masses do not depend on the vev alignment.

Since the mass matrices can be diagonalized analytically, 
we write the quark rotation matrices explicitly, insert them back in the Yukawa lagrangian,
and, for a each choice of the vev alignment, establish how physical scalars interact with quark pairs.  
Taking the vev alignment $B_1$ as a reference,
we find the following interaction terms for the DM candidates $h$ and $a$:
\begin{equation}
	-{\cal L}_Y \supset \frac{m_b}{v}\, \bar b_L \left[(d_R + s_R) h + i (d_R - s_R) a \right] 
	+ \frac{m_t}{v}\, \bar t_L \left[(u_R + c_R) h - i (u_R - c_R) a \right] 
	+ h.c.\label{h-a-quark}
\end{equation}
We arrive at a very concrete conclusion:
the scalars $h$ and $a$, which we considered as DM candidates in the previous section,
cannot be stable as they unavoidable decay to quark pairs with unsuppressed couplings.

We checked that this key result remains valid for all other vev alignments as well as for all 
the irrep choices that can lead to $\Sigma(36)$-symmetric Yukawa sectors, even though
the patterns of the Yukawa matrices and the Higgs-quark couplings can be distinct.
For completeness, in Appendix~\ref{appendix-Yukawa-exact} we provide more details 
for each case. 

In summary, when building an exactly $\Sigma(36)$-invariant Yukawa sector,
there is no way to avoid tree-level decays of the anticipated DM candidates to quark pairs.
These scalars are intrinsically unstable and cannot play the role of dark matter.
The main obstacle is that quarks also carry the conserved quantum numbers with respect 
to the same group $S_3$ that was used to stabilize the scalars $h$ and $a$.
Thus, these scalars are no longer forbidden to decay into quark pairs.

\subsection{A realistic model: $S_3$-symmetric Yukawa sector and softly broken $\Sigma(36)$}\label{section-Yukawa-S3}

Since the exact $\Sigma(36)$ in the Yukawa sector leads not only to the pathological quark sector
but also to unavoidable decays of the anticipated DM candidates,
we need to relax our assumptions. 
We can construct a realistic model using the same $\Sigma(36)$-invariant scalar potential \eqref{Vexact} 
and complementing it with a Yukawa sector that is invariant not under the full $\Sigma(36)$
but under a $S_3$ subgroup.
Moreover, we require this $S_3$ to be exactly the same residual $S_3$ that 
remains unbroken at the minimum.

The specific example we consider is the vev alignment $B_1 = (1, 0, 0)$,
with the residual symmetry group $S_3$ given in Eq.~\eqref{residual-B1}.
The doublet $\phi_1$ is now the trivial singlet of $S_3$, while $\phi_2$ and $\phi_3$ transform
as a doublet irreducible representation.
In the Yukawa sector, we assume that all quarks transform trivially under $S_3$, 
so that they couple only to the first doublet with the SM Yukawa matrices. 
In this way, the doublets $\phi_2$ and $\phi_3$ become truly inert,
and their lightest states are indeed stable.

Of course, the full $\Sigma(36)$ is no longer a conserved symmetry group of the entire lagrangian.
As a result, $\Sigma(36)$-breaking terms will leak, via quark loops, from the Yukawa sector back into the scalar sector.
Instead of calculating these terms, 
we take a different strategy and add soft breaking terms to the scalar potential 
that violate the full $\Sigma(36)$ but preserve $S_3$.
Various soft breaking options for the $\Sigma(36)$ 3HDM were discussed 
in \cite{deMedeirosVarzielas:2021zqs,Yang:2024bys}. If we fix $v$ and $m_h$, 
then there is only one type of $S_3$-preserving soft breaking terms,
\begin{equation}
	V_{\scr{soft}} = \mu^2(\phi_2^\dagger\phi_2 + \phi_3^\dagger\phi_3)\,,\label{soft}
\end{equation}
which must be added to the potential $V_0$ given in Eq.~\eqref{Vexact}.
The requirement that the vev alignment $B_1$ is the global minimum corresponds to $\mu^2 > 0$.
Since the symmetry group $S_3$ is unbroken,
we still have the two DM candidates $h$ and $a$ with the bosonic interactions
coming from the scalar potential and the kinetic terms.
Compared to Eqs.~\eqref{B1-spectrum-charged}--\eqref{B1-spectrum-HA}, the soft breaking 
term leads to a uniform shift 
of the masses of all the scalars from the second and third doublets
keeping the physical scalar doublet degenerate:
\begin{eqnarray}
	(H_2^+, H_3^+): \quad m_{H^+}^2 = \mu^2 + \frac{\lambda_2 v^2}{2}\,,\qquad 
	(h, a): \quad m_{h}^2 = \mu^2 + \frac{\lambda_3 v^2}{2}\,,\qquad
	(H, A): \quad m_{H}^2 = \mu^2 + \frac{3 \lambda_3 v^2}{2}\,.\quad \label{B1-spectrum-soft}
\end{eqnarray}
We then choose $m_h$, $m_{H^+} > m_h$, and $\mu^2$ as the three independent free parameters\footnote{Another possible choice
of the three free parameters is $m_h$, $m_H$, and $m_{H^+} > m_h$.}
and express the quartic couplings $\lambda_2$ and $\lambda_3$ by inverting the first two relations
from Eq.~\eqref{B1-spectrum-soft}, while the heavy inert Higgs mass is given by 
$m_H^2 = 3 m_h^2 - 2\mu^2$.
Note that all three quantities $\mu^2$, $\lambda_2$, $\lambda_3$ must be positive for the global minimum to reside at $B_1$.
Also, due to the present of $\mu^2$, the model now has a well-defined decoupling limit.

We still call this model the $\Sigma(36)$-based DM model.
Although the exact symmetry group is $S_3$, the quartic potential is the same as in the $\Sigma(36)$-symmetric 3HDM,
which constrains in a significant way the types of interactions and imposes correlations among the couplings.
This model ``inherits'' certain features from the $\Sigma(36)$-symmetric 3HDM \cite{deMedeirosVarzielas:2021zqs},
and this is why we distinguish it from a generic $S_3$-based 3HDM DM models \cite{Khater:2021wcx,Kuncinas:2022whn}.


\section{Dark matter properties in the $\Sigma(36)$-based 3HDM}\label{section-DM}

\subsection{Constraints on the model and comparison with the IDM}

We implemented the $\Sigma(36)$-based DM model just described in micrOMEGAs \cite{Alguero:2023zol}, 
which allowed us to explore the relic density and direct detection (DD) signals
as functions of the free parameters of the model.
Note that the model features an exact scalar alignment limit, which allows us to evade most of the phenomenological
constrains on the SM-like Higgs measurements.
We must, however, take into account the important constraint on the model coming from the SM-like Higgs decay to the DM candidates, 
which exists for sufficiently light $h$ and $a$.
The negative results of the LHC searches for an invisible Higgs decay \cite{ATLAS:2023tkt} lead 
to the upper limit on its branching ratio, $B_{\scr{inv}} < 0.107$,
which translates into $\Gamma(\hSM \to \mbox{inv.}) < 0.42$~MeV.

To discuss the resulting constraints on the free parameters of our model,
we find it instructive to compare it with the IDM in the notation of \cite{Ilnicka:2015jba,Belyaev:2016lok,Kalinowski:2018ylg}.
The scalar potential of the IDM contains seven free parameters. After fixing $v$ and $\mSM$,
we still have five real parameters to adjust, which affect the IDM phenomenology.
Among them, the key role is played by the trilinear $\hSM$-DM-DM
coupling divided by $v$, which is traditionally labeled as $\lambda_{345}$. This coefficient sets the magnitude
of several important processes, such as DM annihilation through the $s$-channel Higgs resonance and the invisible Higgs decay to light DM candidates.
It also contributes to the quartic vertices $hhW^+_LW^-_L$ and $hhZ_LZ_L$ to be discussed below.
The key feature is that, within the IDM, this all-important $\lambda_{345}$ is an independent parameter, 
not directly related to the scalar masses. 
It can be chosen sufficiently small, $|\lambda_{345}| \lsim 0.04$, 
so that, for the light DM candidates, the Higgs invisible decay width stays suppressed.

However, in our model, the coefficient in the vertices $\hSM hh$ and $\hSM aa$ is not an independent parameter but is expressed
via masses of the $\hSM$ and the SM candidates:
\begin{equation}
	\hSM hh,\ \hSM aa: \quad \bar\lambda v \equiv (2\lambda_1 + \lambda_3)v = \frac{\mSM^2 + m_H^2 - m_h^2}{v}
	= \frac{\mSM^2 + 2 m_h^2 - 2\mu^2}{v}\,,\label{hh-aa}
\end{equation}
which is never too small. It is at least as large as $v/4$ and grows further as the inert sector mass splitting increases.
This coefficient leads to a huge invisible decay width of the SM-like Higgs 
\begin{equation}
	\Gamma(\hSM \to hh, aa) = \frac{(\mSM^2 + 2 m_h^2 - 2\mu^2)^2}{16\pi \mSM v^2} \beta\,, \quad
	\mbox{where}\ \beta = \sqrt{1 - \frac{4 m_h^2}{\mSM^2}}\,.\label{invisible}
\end{equation}
For $m_h^2 = \mu^2$, it yields $640\,\mbox{MeV}\cdot \beta$
and exceeds the experimental upper limit by orders of magnitude.
Even the nominal threshold $\mSM = 2 m_h$ is excluded: a slightly off-shell Higgs boson 
with the invariant mass exceeding the nominal value $\mSM$ by a few $\Gamma_{\scr{SM}}$ would already generate an unacceptably large signal.
This analysis implies that in our model, unlike in the IDM, the light DM region $m_h \le \mSM/2$ is altogether ruled out.

\begin{figure}[!ht]
	\centering
	\includegraphics[height=6cm]{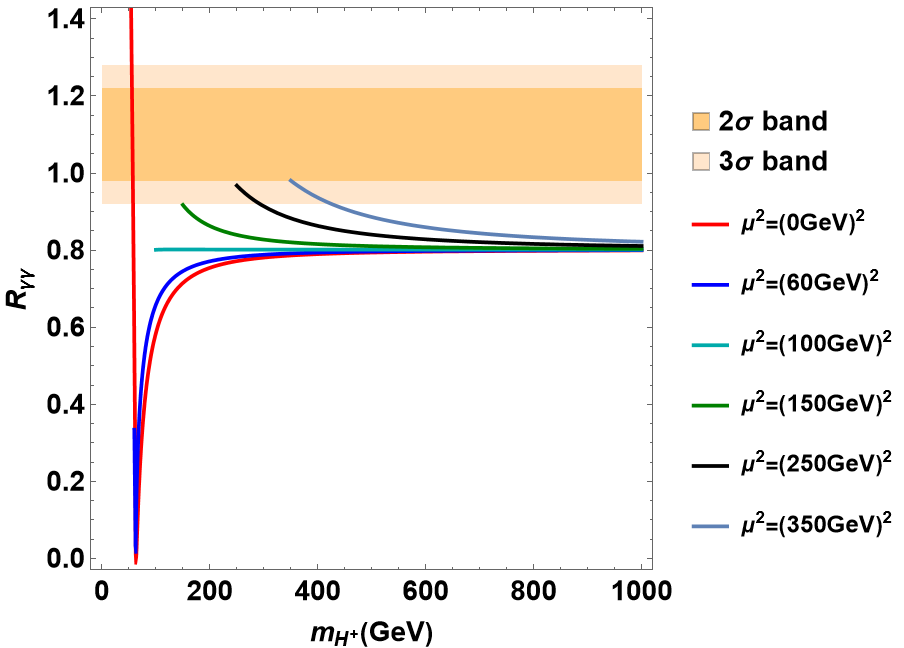}
	\caption{The ratio $R_{\gamma\gamma}$ defined in Eq.~\eqref{Rgamgam} as a function of he charged Higgs mass for several values of $\mu^2$ compared with 
		the LHC results for the $2\sigma$ and $3\sigma$ bands around the central value.}
	\label{fig-Rgamgam}
\end{figure}

The presence of inert scalars also modifies the loop-induced decays of the SM-like Higgs, in particular, $\hSM \to \gamma\gamma$. 
Let us denote the modification of this decay width with respect to the SM prediction as
\begin{equation}
	R_{\gamma\gamma} = \frac{\Gamma(\hSM \to \gamma\gamma)}{\Gamma(\hSM \to \gamma\gamma)^{\mbox{\scriptsize SM}}}\,.\label{Rgamgam}
\end{equation}
The expression for this decay width can be recovered from the generic 2HDM result \cite{Djouadi:2005gj}:
\begin{eqnarray}
	\Gamma(\hSM \to \gamma\gamma)&=&\frac{G_F\alpha^2 \mSM^3}{128\sqrt{2}\pi^3}\bigg | 
	\sum_f N_c Q_f^2 A_{1/2}(\tau_f) + A_1(\tau_W)
	+2\times \frac{\mSM^2 + 2 m_{H^+}^2 - 2\mu^2}{2 m_{H^+}^2}A_0(\tau_+)\bigg |^2,\label{gamgam}
\end{eqnarray}
while for the SM expression we simply omit the last contribution.
The factor 2 in the last term shows that both $H_2^+$ and $H_3^+$ give equal contributions.
Following \cite{Djouadi:2005gj}, for a particle with mass $m_i$, we define the mass ratio as $\tau_i = \mSM^2/(4m_i^2)$
and use the formfactors 
\begin{eqnarray}
	A_0(\tau) = -\frac{\tau-f(\tau)}{\tau^{2}},\quad
	A_{1/2}(\tau) = 2\frac{\tau+(\tau-1)f(\tau)}{\tau^{2}},\quad
	A_1(\tau) = -\frac{2\tau^2+3\tau+3(2\tau-1)f(\tau)}{\tau^{2}}\,,
\end{eqnarray}
where
\begin{equation}
	f(\tau)= \left\{\arcsin^2\sqrt{\tau} \ \ \mbox{for $\tau\le 1$}, \quad
		-\frac{1}{4}\left[\log\left(\frac{\displaystyle 1+\sqrt{1-\tau^{-1}}}{ \displaystyle  1-\sqrt{1-\tau^{-1}}}\right)-i\pi\right]^2  
		\ \mbox{for $\tau > 1$} \right\}.
\end{equation}
The structure of the charged Higgs contribution closely resembles the IDM counterpart \cite{Cao:2007rm,Arhrib:2012ia,Swiezewska:2012eh,Krawczyk:2013jta}.
However, unlike in the IDM, the sign of the coefficient is fixed, which leads to $R_{\gamma\gamma} < 1$ for a heavy charged Higgs. 

The coefficient in the last term of Eq.~\eqref{gamgam} makes it clear that, without the soft breaking terms, 
the contribution of the charged Higgses is not negligible. 
We find that, for $\mu^2 = 0$ and heavy charged Higgses, $R_{\gamma\gamma}$ is as low as $0.8$, which is already ruled out 
by the LHC measurements \cite{ParticleDataGroup:2024cfk}: $R^{\scriptsize exp.}_{\gamma\gamma} = 1.10 \pm 0.06$.
However, a non-zero $\mu^2$ reduces the tension, provided the charged Higgs mass stays close to $\mu$.
In Fig.~\ref{fig-Rgamgam}, we show the behavior of $R_{\gamma\gamma}$ as a function of $m_{H^+}^2$
for several values of $\mu^2$.
As we can see, the $R_{\gamma\gamma}$ curves can reach $0.92$, which lies within $3\sigma$ of the central experimental value,
only for  $\mu^2 > (150\,\mbox{GeV})^2$, and can exceed $0.98$ ($2\sigma$ band) only for $\mu^2 > (350\,\mbox{GeV})^2$.

We arrive at the conclusion that the charged Higgs contributions to $\hSM \to \gamma\gamma$ are very sizable
and, in the absence of any additional contribution to compensate their effect,
rule out the $\Sigma(36)$ symmetric scalar sector. 
The only way to bring $\hSM \to \gamma\gamma$ back within experimental limits is to assume a sufficiently large soft breaking term.

\subsection{DM properties: no soft breaking}

\begin{figure}[!htb]
	\centering
	\includegraphics[height=6cm]{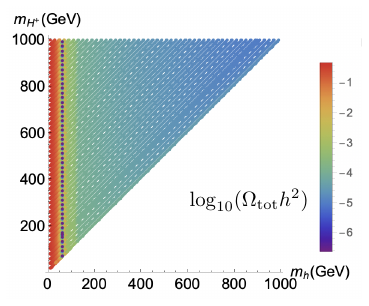}
	\hfill
	\includegraphics[height=5.5cm]{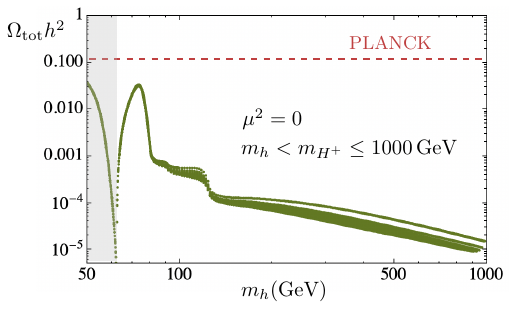}
	\caption{Combined relic density of $h$ and $a$ in the 3HDM with an exact $\Sigma(36)$. 
		Left: the relic density encoded in color on the parameter plane $(m_h, m_{H^+})$.
	Right: the relic density as a function of $m_h$ for several values of $m_{H^+}$.}
	\label{fig-exact-relic}
\end{figure}

Keeping the above results in mind, we turn now to the study of the relic density as a function of the free parameters.
We begin the presentation of our results 
with the case of the $\Sigma(36)$-symmetric scalar sector, that is, the model without soft breaking terms, $\mu^2 = 0$.
Although the measurements of the two photon decay width effectively rule out this scenario,
we still want to see whether additional problems arise from the DM sector alone.

When exploring the relic density evolution, we always observe that $h$ and $a$ give equal contributions.
This is of course expected due to the symmetry that links $h$ and $a$, the two components of the $S_3$ doublet.
Thus, when showing the values for $\Omega h^2$, we will always present the sum of their contributions.

In Fig.~\ref{fig-exact-relic}, we show the resulting relic density $\Omega h^2$ as a function of the two remaining free parameters
$m_h$ and $m_{H^+}$. The left plot shows the results of a general scan, with the relic density encoded in color,
while the left plot reveals finer details of $\Omega h^2$ as a function of $m_h$ for several representative values of $m_{H^+}$
that go up to 1 TeV.
The shaded region in the right plot corresponds to $m_h < \mSM/2$ and is excluded by the invisible decay width.

A salient feature of these plots is that, for $m_h > 45$~GeV, the relic density is always
below the Planck result \cite{Planck:2018vyg}: $(\Omega h^2)_{\scr{Planck}} = 0.1200 \pm 0.0012$.
This is due to the rather large annihilation, coannihilation, and semi-annihilation cross sections.
The only region where the predicted relic density matches the Planck result is the narrow band around $m_h \approx 40$~GeV,
but this low mass region is already excluded by the invisible Higgs decay constraint.

The low relic density in the high-mass region can also be understood by comparing our model with the IDM.
In the IDM with heavy DM candidates, annihilation mainly proceeds into the longitudinally polarized $W^+W^-$ and $ZZ$ pairs
\cite{Belyaev:2016lok}, which are dominated by the quartic vertices 
\begin{equation}
	hhW^+_LW^-_L\mbox{(IDM):}\quad \lambda_{345} + \frac{2(m_H^2-m_h^2)}{v^2}\,, \qquad
	hhZ_LZ_L\mbox{(IDM):}\quad \lambda_{345} + \frac{2(m_{H^+}^2-m_h^2)}{v^2}\,.\label{IDM-quartic}
\end{equation} 
As mentioned earlier, $\lambda_{345}$ is an independent parameter and can be chosen small. 
If the mass splitting in the inert sector is also small, the annihilation cross section is suppressed, 
and the relic density can match the observed value.
This is the mechanism behind the high-mass region of the IDM, which still remains viable.

However, in the $\Sigma(36)$ model, these couplings are
\begin{equation}
	hhZ_LZ_L: \quad 2\lambda_1 + 3\lambda_3 = \bar\lambda + \frac{2(m_H^2-m_h^2)}{v^2}\,,\qquad
	hhW^+_LW^-_L: \quad 2\lambda_1 + \lambda_2 = \bar\lambda + \frac{2(m_{H^+}^2-m_h^2)}{v^2}\,,\label{hhVV}
\end{equation}
where $\bar\lambda$ is given in Eq.~\eqref{hh-aa}.
Both expressions are never small. The minimal value for the $hhZ_LZ_L$ coupling in the exactly $\Sigma(36)$-symmetric scalar sector 
is $5\mSM^2/2v^2 \approx 0.6$ and further grows approximately as $m_h^2$.
This explains why, for large DM masses, we always obtain the relic density significantly below the observed value.

We conclude that the model can account only for a small fraction of total dark matter.
Although it still leaves open the question of the origin of the dominant DM component, 
this sub-dominant DM scenario is, in principle, viable.

Next, we confront this model with the direct detection limits.
Let us denote by $\xi_\Omega = \Omega/\Omega_{\scr{Planck}}$ the fraction of the total relic density constituted by $h$ and $a$. 
Then, we can compute the DD signal expected in this case, taking into account the sub-dominant DM scenario we work in,
and compare it with the experimental upper limits on the spin-independent cross section $\sigma_{SI}$ established by the DD searches.
Instead of up-scaling the published upper limits on $\sigma_{\scr{SI}}$ for such a comparison, 
we downscale our predicted cross section by defining 
\begin{equation}
\hat\sigma_{SI} = \sigma_{SI}\cdot \xi_\Omega\,,\label{hat-sigma}
\end{equation}
and directly compare it with the published results.

\begin{figure}[!htb]
	\centering
	\includegraphics[height=6cm]{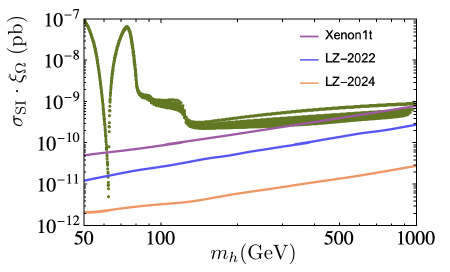}
	\caption{The rescaled cross section for the direct detection signal $\hat\sigma_{SI} = \sigma_{SI}\cdot \xi_\Omega$ obtained in our model,
		compared with the upper limits from recent DD experiments.}
	\label{fig-exact-DD}
\end{figure}

In Fig.~\ref{fig-exact-DD} we present this comparison. Even with the scaling down taken into account, the model predicts rather large DD signals. 
This is unavoidable because the DM-Higgs interaction is governed by the same coupling $\bar\lambda$ as in Eq.~\eqref{hh-aa}.
Still, there remains a sizable part of the random scan which passes the limits obtained by the Xenon1T experiment.
However, these predictions are in a clear conflict with the upper limits announced by the LZ experiment 
both in 2022 \cite{LZ:2022lsv} and especially in 2024 \cite{LZCollaboration:2024lux}, definitely ruling out the model we consider.

The bottom line is: with the latest LZ-2024 result, the scenario featuring two DM candidates arising 
within the 3HDM with an exact $\Sigma(36)$ symmetry in the scalar sector is ruled out.

\subsection{DM properties with soft breaking terms}\label{subsection-DM-soft}

\begin{figure}[h]
	\centering
	\includegraphics[height=5cm]{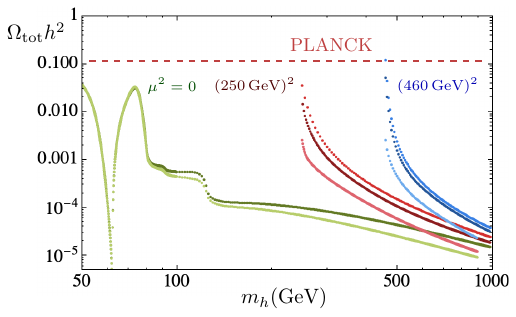}
	\hfill
	\includegraphics[height=5cm]{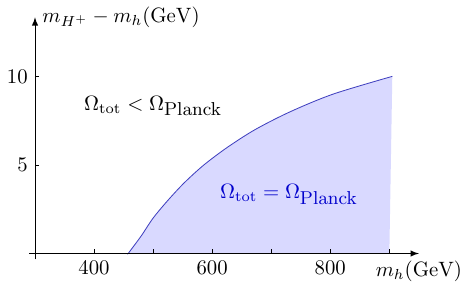}
	\caption{Left: The total relic density in the 3HDM for three values of the soft breaking parameter $\mu^2$ and
		several options for $m_{H^+} - m_h$, see the main text. 
		Right: The region in the $(m_{H^+} - m_h, \,m_h)$ parameter space 
		in which the relic density computed with a suitable $\mu^2$ can match the Planck results.}
	\label{fig-soft-relic}
\end{figure}

%

Next, we introduce the soft breaking terms \eqref{soft} and explore the consequences for the DM observables.
In Fig.~\ref{fig-soft-relic}, left, we show how the DM relic density depends on $\mu^2 > 0$.
The light and dark shaded green points correspond to the exactly $\Sigma(36)$-symmetric scalar sector 
with $m_{H^+} - m_h = 10\,\mbox{GeV}$ and $100\,\mbox{GeV}$;
they agree with Fig.~\ref{fig-exact-relic}, right.
The red and blue families of points correspond to $\mu^2 = (250\,\mbox{GeV})^2$ and $(460\,\mbox{GeV})^2$,
respectively. In both cases, the three sequences of points, from top to bottom, refer to 
$m_{H^+} - m_h = 0.1\,\mbox{GeV}$, 10~GeV, and 100~GeV. 
As expected, if $m_h$ is fixed but $\mu$ increases, the annihilation cross section goes down, 
and as a result the relic density rises.

Fig.~\ref{fig-soft-relic}, left, indicates that, for $\mu^2 > (460\,\mbox{GeV})^2$ and with a small mass splitting,
a region of parameters opens up, in which the calculated relic density matches the Planck measurements.
For a better visualization, we show this region in Fig.~\ref{fig-soft-relic}, right, on the 
plane of the mass splitting $m_{H^+} - m_h$ vs. $m_h$.
For any choice of masses $m_{H^+}$ and $m_h$ that fall inside the blue region, 
it is possible to find a suitable $\mu^2$ that leads to the total relic density
that matches the Planck result. For masses $m_{H^+}$ and $m_h$ outside this region,
one always obtains subdominant scalar DM for any choice of $\mu^2$.
It is interesting to note that this borderline value of $\mu$ 
is very similar to the value of the DM mass in the IDM, $m_{DM} \sim 500$ GeV, above which the so-called heavy mass region
with a matching relic density opens up \cite{Arhrib:2013ela,Ilnicka:2015jba,Belyaev:2016lok}.

\begin{figure}[!htb]
	\centering
	\includegraphics[height=5cm]{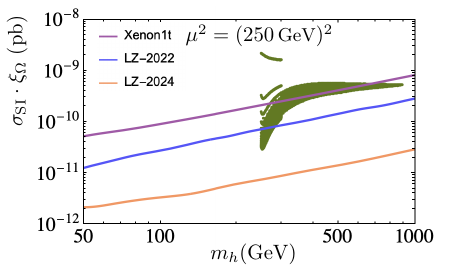}
	\hfill
	\includegraphics[height=5cm]{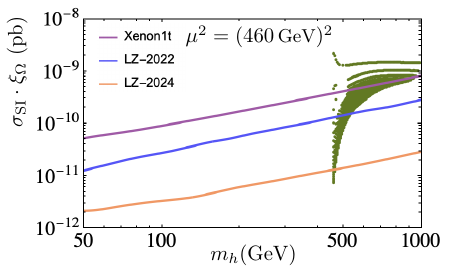}
	\caption{The rescaled cross section for the direct detection signal $\hat\sigma_{SI} = \sigma_{SI}\cdot \xi_\Omega$ for 
		$\mu^2 = (250\,\mbox{GeV})^2$ and $(460\,\mbox{GeV})^2$,
		compared with the experimental upper limits.}
	\label{fig-soft-DD}
\end{figure}

In Fig.~\ref{fig-soft-DD}, we show how the rescaled cross section for the direct detection signal $\hat\sigma_{SI}$ 
changes as $\mu^2$ grows. We show the predictions for the same soft breaking parameter values:
$\mu^2 = (250\,\mbox{GeV})^2$ and $(460\,\mbox{GeV})^2$.
For DM mass well above $\mu$, the rescaled DD signals still hover above the upper limits from LZ,
similarly to the case $\mu^2=0$, and are not significantly suppressed with $\mu^2$ variation.
This behavior can be understood by noticing once more that the same $\bar\lambda$, which is never too small, 
both sets the DM scattering process as in Eq.~\eqref{hh-aa} and drives the annihilation cross section, Eq.~\eqref{hhVV}.
If we fix a large $m_h$ and increase $\mu^2$, the DD cross section goes down, but the relic density fraction goes up,
and the two effects approximately compensate each other in the rescaled DD signal.

The only way to keep DD cross section low and, at the same time, further suppress relic density
is to keep $\bar\lambda$ small and increase the charged Higgs mass.
This is what we observe for $m_h$ very close to $\mu$. In particular, we find that the model with 
$\mu^2 = (460\,\mbox{GeV})^2$ is, in principle, capable of bringing the DD signal
below not only LZ-2022 but also the latest LZ-2024 upper limits \cite{LZCollaboration:2024lux}.
However, this is done at the expense of a large $\lambda_2$;
we checked that the lowest lying points in Fig.~\eqref{fig-soft-DD}, right, correspond to 
$m_{H^+} > 720$~GeV, which translates into $\lambda_2$ above 10.
Also, the large mass splitting between $H, A$ and $H_{2,3}^+$, which arises in this case,
drives the electroweak precision observables far beyond the experimental limits.

We conclude that the tight correlations arising in our model even in the softly broken regime, 
once more, lead the predictions to severe clashes with theoretical and experimental constraints.
It may prove useful to extend the parameter $\mu^2$ to even larger values in a quest for a fully viable DM model.
But we believe that the present study clearly indicates the persistent problems
that arise in this and other similar DM models based on a multi-Higgs-doublet with very large finite symmetry groups.

\section{Discussion and conclusions}

In this paper we addressed the question whether scalar DM candidates stabilized by a non-abelian residual group 
is a viable option within multi-Higgs-doublet models. 
A conserved non-abelian group not only protects the multiplet of DM candidates against decay 
but also gives rise to additional DM evolution channels. 
In contrast to previous works, in which a non-abelian group protected dark sector, by construction, 
was nearly decoupled from the SM fields, 
we considered a multiplet of DM candidates emerging from additional inert electroweak scalar doublets 
that enter the same Higgs potential that leads to the SM-like Higgs boson. 
As the benchmark example, we used the $\Sigma(36)$-based 3HDM \cite{Ivanov:2012ry,Ivanov:2012fp}
and took the advantage of its property that, for any choice of its free parameters,
the global minimum of the potential always preserves a non-abelian subgroup $S_3 \subset \Sigma(36)$ 
and therefore always stabilizes a pair of mass degenerate DM candidates \cite{Ivanov:2014doa}.
 
	Due to very tight symmetry-based relations between the SM-like Higgs, the quark sector, and the dark sector,
we found that this idea repeatedly runs into conflict with observations.
\begin{itemize}
	\item 
The 3HDM Yukawa sector invariant under the exact $\Sigma(36)$
not only leads to non-physical quark properties but also destabilizes all the scalars,
rendering a realistic quark sector incompatible with DM candidates.
To avoid this problem, we relaxed our assumptions and built the Yukawa sector invariant not under the full $\Sigma(36)$
but under $S_3$, the same subgroup that remains unbroken after minimization.
	\item
The extra charged Higgses significantly reduce the $\hSM \to \gamma\gamma$ decay width, which conflicts the LHC measurements.
As the $\hSM$ coupling with DM is never suppressed, the unsuppressed invisible Higgs decay rules out this model
for $m_h < \mSM/2$.
	\item
If the scalar sector respects the full $\Sigma(36)$ symmetry, the scalar DM is unavoidably subdominant and, 
at the same time, enters in conflict with the direct detection results from the LZ experiment \cite{LZCollaboration:2024lux}. 
The bottom line is that the model with an exactly $\Sigma(36)$-symmetric scalar sector 
complemented with $S_3$-invariant Yukawa interactions is ruled out.
	\item
The soft breaking terms governed by the single new free parameter $\mu^2$ can relax the tension.
These soft breaking terms violate $\Sigma(36)$ but preserve the same $S_3$ that remains as a residual symmetry group at the minimum,
thus, supporting the same pair of DM candidates stabilized by the residual non-abelian group.
Still, the quartic potential inherits many features of the exact $\Sigma(36)$ 3HDM and remains very constrained. 
A moderately large $\mu > 350$ GeV can bring the $\hSM \to \gamma\gamma$ within the experimental limits.
For $\mu > 460$ GeV, a high-mass, compressed-spectrum region opens up, in which the relic density predictions
can match the Planck result. However, the DD signals in this region are way too high, 
above not only LZ \cite{LZ:2022lsv,LZCollaboration:2024lux} but even Xenon-1T \cite{XENON:2018voc} results.
	\item
Alternatively, there is a region for $\mu > 460$ GeV with a very sub-dominant DM scenario, in which the DD signal
dives below the latest LZ limits \cite{LZCollaboration:2024lux}. However, it requires a large charged vs. neutral Higgs mass splitting,
and an extreme value of $\lambda_2$, violating experimental and theoretical constraints.
\end{itemize}
The origin of these repeated conflicts is clear.
Our starting idea of a large finite global symmetry group strongly constraints the shape of the scalar sector.
Looking back at the Higgs potential \eqref{Vexact}, we see that it is the term 
$\lambda_1(\phi_1^\dagger \phi_1+ \phi_2^\dagger \phi_2+\phi_3^\dagger \phi_3)^2$
that forces the effective DM-SM coupling $\bar\lambda$ in Eq.~\eqref{hh-aa} to always remain
unsuppressed. It is this unsuppressed coupling that drives the large invisible Higgs decay for light DM,
the large contribution of the charged Higgs loops to $\hSM \to \gamma\gamma$ for heavy DM,
as well as the large DD signal.
Thus, the large DM effects are tightly connected with our starting assumption of a large symmetry group.

It may happen that extending our study to $\mu^2$ in excess of $1\,\mbox{TeV}^2$ would reveal a model that would be 
technically within the existing limits. But judging from the present experience, it will always be at the brink of exclusion
and never be a natural candidate.
But the main goal of this work was not to construct yet another DM model but 
to explore the limits of what multi-Higgs-doublet models can in principle accommodate.
We met this objective: through a concrete example, we learned several lessons 
that further highlight the conflicts between symmetry assumptions,
the LHC results, and DM properties that arise if we try to derive a DM multiplet stabilized by a non-abelian group
from the multi-Higgs-doublet model potential.
We believe that these lessons are of interest to the community and help clarify what can and what cannot 
be achieved with several Higgs doublets alone, without additional New Physics assumptions.

\section*{Acknowledgments}

This work is supported by Guangdong Natural Science Foundation (project No.~2024A1515012789).
The work of R.B. and J.P.S. is supported in part by the Funda\c{c}\~{a}o para a Ci\^{e}ncia e a Tecnologia (FCT), under Contracts 2024.01362.CERN, UIDB/00777/2020, and UIDP/00777/2020; these projects are partially funded through POCTI (FEDER), COMPETE, QREN, and the EU. The work of R. Boto is also supported by FCT with the PhD grant PRT/BD/152268/2021.


\appendix

\section{$SU(3)$ vs. $PSU(3)$}\label{appendix-technical}

When we discuss three-family global symmetries either of the Higgs doublets or of fermion generations, 
we usually think of a symmetry group $G \subset SU(3)$. 
However, $SU(3)$ contains a non-trivial center: $Z(SU(3)) \simeq \Z_3 = 
\{\id_3, \, \omega\cdot \id_3, \, \omega^2\cdot \id_3\}$.
Of course, this center is always a symmetry of the entire lagrangian of any 3HDM, just because it is a very particular 
subgroup of the hypercharge gauge transformations $U(1)_Y$.
Since we are looking for symmetry groups $G$ {\em in addition} to the electroweak gauge group, 
we should not include this $\Z_3$ in our search for global symmetry groups.
This can be done by considering the factor group $SU(3)/\Z_3$, which is called $PSU(3)$, the projective $SU(3)$.
Thus, if we aim to establish all distinct symmetry-based situations in the 3HDM, we should look for subgroups 
of $PSU(3)$, not $SU(3)$.

Let us show how it works for the group $\Delta(27)$ mentioned above.
The two generators in Eq.~\eqref{Delta27-generators} do not commute inside $SU(3)$, as their commutator gives 
\begin{equation}
	[a,b] = aba^{-1}b^{-1} = \omega^2 \cdot \id_3\,.
\end{equation}
However, if we now view $a$ and $b$ as representative elements of the cosets $a\Z_3$ and $b\Z_3$, which belong to $PSU(3)$,
then they do commute: their commutator is just the coset $\Z_3$ itself, the trivial element of $PSU(3)$.
So, by applying the homomorphism $f: SU(3) \to PSU(3)$ to the group $\Delta(27) \subset SU(3)$,
we obtain the abelian group $\Delta(27)/\Z_3 = \Z_3 \times \Z_3 \subset PSU(3)$. 
Similarly, if we start with $\Delta(54) = \Delta(54)\rtimes\Z_2 \subset SU(3)$, 
we arrive at $\Delta(27)/\Z_3 = (\Z_3 \times \Z_3)\rtimes \Z_2 \subset PSU(3)$.
This is why the classification of the finite symmetry groups of the 3HDM scalar sector \cite{Ivanov:2011ae,Ivanov:2012ry,Ivanov:2012fp},
dealt with $(\Z_3 \times \Z_3)\rtimes \Z_2$, not $\Delta(54)$.

However, working directly inside $PSU(3)$ is less convenient.
We are used to exploring complex spaces, not projective spaces,
and to dealing with matrices as in Eq.~\eqref{Delta27-generators}, not cosets of matrices.
Moreover, we have much information on unitary irreducible representations (irreps) of finite groups, not projective irreps.
Therefore, in practical calculations, it is advisable to work inside $SU(3)$---remembering about the center
when appropriate.

The group $\Sigma(36)$ of order 36 was identified in 
\cite{Ivanov:2011ae,Ivanov:2012ry,Ivanov:2012fp} as the largest finite symmetry group of the 3HDM scalar sector.
This is a subgroup of $PSU(3)$, and it can be written as $(\Z_3 \times \Z_3)\rtimes \Z_4$, 
with the already familiar $\Z_3 \times \Z_3$ ``core'' and the new transformation which acts on it by automorphisms. 
Back within $SU(3)$, this group is denoted as $\Sigma(36\varphi)$, or $\Sigma(36\times 3)$, 
see e.g. \cite{Grimus:2010ak,Hagedorn:2013nra}, and is of order 108.

\section{Representations of the group $\Sigma(36\varphi)$}\label{appendix-group}

Irreducible representations of the group $\Sigma(36\varphi)$ together with the character table 
were listed in \cite{Grimus:2010ak,Hagedorn:2013nra}.
They include four 1D irreps $\bm{1}^{(p)}$, which differ by the representing value of $\rho_1(d) = i^p$,
four complex 3D irreps $\bm{3}^{(p)}$, whose $\rho_3(a)$ and $\rho_3(b)$ are given in Eq.~\eqref{Delta27-generators}
and $\rho_3(d)$ is the same as in Eq.~\eqref{Sigma36-generators} with an extra factor $i^{p}$,
four conjugate triplets $(\bm{3}^{(p)})^*$, and finally two real 4D irreps.
As in the main text, we shorten $\bm{3}^{(0)}$ to just $\bm{3}$.
The sum of the squares of the irrep dimensions is $\sum_i d_i^2 = 4\times 1 + 8\times 3^2 + 2\times 4^2 = 108$,
which is the order of $\Sigma(36\varphi)$.

The decomposition of various products of irreps were derived in \cite{Grimus:2010ak}, Appendix C.
We are interested in products of $\bm{3}$, which is how the Higgs doublets transform,
with $\bm{3}^{(p)}$ or $(\bm{3}^{(p)})^*$.
First,
\begin{equation}
	\bm{3} \otimes \bm{3}^* = \bm{1}^{(0)} \oplus \bm{4} \oplus \bm{4}'\,.\label{CG-3030*}
\end{equation}
If $a \sim \bm{3}$ and $b^* \sim \bm{3}^*$, then the trivial singlet is obtained as
\begin{equation}
	[ab^*]_1 = a_1 b_1^* + a_2 b_2^* + a_3 b_3^*\,. 
\end{equation}
Eq.~\eqref{CG-3030*} can be generalized to other triplets-antitriplet products as
\begin{equation}
	\bm{3}^{(p)} \otimes (\bm{3}^{(p')})^* = \bm{1}^{(p-p')} \oplus \bm{4} \oplus \bm{4}'\,,\label{CG-3p3p*}
\end{equation}
where $p-p'$ is understood as taken mod $4$.

Next, the product of two triplets is decomposed as 
\begin{equation}
	\bm{3} \otimes \bm{3} = \bm{3}^* \oplus (\bm{3}^{(1)})^* \oplus (\bm{3}^{(3)})^*\,.\label{CG-3030}
\end{equation}
If $a \sim \bm{3}$ and $b \sim \bm{3}$, then the explicit expression for the antitriplet $\bm{3}^*$, which is antisymmetric under $a\leftrightarrow b$, 
is inherited from the $SU(3)$ irrep products:
\begin{equation}
	\frac{1}{\sqrt{2}}\epsilon_{ijk}a_jb_k = \frac{1}{\sqrt{2}}\triplet{a_2 b_3 - a_3 b_2}{a_3 b_1 - a_1 b_3}{a_1 b_2 - a_2 b_1} \sim \bm{3}^*\,.
	\label{CG3030-30*}
\end{equation}
The six-dimensional symmetric subspace splits into the two invariant subspaces: 
\begin{equation}
	\frac{1}{\sqrt{24}}\triplet{\sqrt{2}\,\tau_- a_1 b_1 - \tau_+(a_2 b_3 + a_3 b_2)}%
							   {\sqrt{2}\,\tau_- a_2 b_2 - \tau_+(a_3 b_1 + a_1 b_3)}%
							   {\sqrt{2}\,\tau_- a_3 b_3 - \tau_+(a_1 b_2 + a_2 b_1)} \sim (\bm{3}^{(1)})^*\,, \quad
	\frac{1}{\sqrt{24}}\triplet{\sqrt{2}\,\tau_+ a_1 b_1 + \tau_-(a_2 b_3 + a_3 b_2)}%
							   {\sqrt{2}\,\tau_+ a_2 b_2 + \tau_-(a_3 b_1 + a_1 b_3)}%
							   {\sqrt{2}\,\tau_+ a_3 b_3 + \tau_-(a_1 b_2 + a_2 b_1)} \sim (\bm{3}^{(3)})^*\,.
	\label{CG3030-3p*}
\end{equation}
Here, we used the shorthand notation
\begin{equation}
	\tau_- = \sqrt{2(3-\sqrt{3})}\,, \quad 	\tau_+ = \sqrt{2(3+\sqrt{3})}\,, \quad \tau_-\tau_+ = \sqrt{24}\,.
\end{equation}
Again, the decomposition Eq.~\eqref{CG-3030} can be generalized to the product of any two triplets:
\begin{equation}
	\bm{3}^{(p)} \otimes \bm{3}^{(p')} = (\bm{3}^{(-p-p')})^* \oplus (\bm{3}^{(1-p-p')})^* \oplus (\bm{3}^{(3-p-p')})^*\,,\label{CG-3p3p}
\end{equation}
where all indices are computed mod $4$. However, the decomposition rules remain the same as in Eqs.~\eqref{CG3030-30*} and \eqref{CG3030-3p*}.


\section{Yukawa sectors for the exact $\Sigma(36)$ symmetry}\label{appendix-Yukawa-exact}

We present here the details of extending the symmetry group $\Sigma(36)$ to the quark Yukawa sector.
With the list of irreps of the group $\Sigma(36\varphi)$ and their product composition rules,
we present in Table~\ref{table-options} the main options 
for the irrep assignments in a $\Sigma(36\varphi)$-invariant quark sector.
In principle, one can also choose different triplets; for example, 
in the second line, $\overline{Q_L}$ could be chosen to transform as $\bm{3}^{(1)}$.
But with a simultaneous relabeling the irrep assignments for $d_R$ and $u_R$, 
this choice would still lead to the same Yukawa structures.
Let us now build the Yukawa matrices for each of the cases listed in Table~\ref{table-options}. 

\begin{table}[!h]
	\centering
	\begin{tabular}[t]{ccccc}
		\toprule
		label & $\overline{Q_L}$ & $\phi$ & $d_R$ & $u_R$\\
		\midrule
		case 1-3-3-3\quad & $\quad\bm{1}\quad$ & $\quad\bm{3}\quad$ & $\quad\bm{3}^*\quad$ & $\quad\bm{3}\quad$ \\
		case 3-3-1-3\quad & $\bm{3}^*$ & $\bm{3}$ & $\bm{1}$ & $\bm{3}^*$ or $(\bm{3}^{(1)})^*$ or $(\bm{3}^{(3)})^*$ \\
		case 3-3-3-1\quad & $\bm{3}$ & $\bm{3}$ & $\bm{3}$ or $\bm{3}^{(1)}$ or $\bm{3}^{(3)}$ & $\bm{1}$ \\
		\bottomrule
	\end{tabular}
	\caption{The main options for $\Sigma(36\varphi)$-invariant Yukawa sector.}
	\label{table-options}
\end{table}

{\bf Case 1-3-3-3}: $Q_L$ are trivial singlets, $d_R \sim \bm{3}^*$, $u_R \sim \bm{3}$, the first line of Table~\ref{table-options}.
The three Yukawa matrices $\Gamma_a$ were given in Eq.~\eqref{case1333-Gamma},
and similar matrices $\Delta_a$ with their own parameters $d_i$ arise for the up sector.
Starting from the mass matrices $(M_d)_{ij} = g_i v_j/\sqrt{2}$ and $(M_u)_{ij} = d_i v^*_j/\sqrt{2}$,
we define, as usual, their hermitean squares $H_d = M_d M_d^\dagger$ and $H_u = M_u M_u^\dagger$: 
\begin{equation}
	(H_d)_{ij} = \frac{v^2}{2}g_i g^*_j\,, \quad 
	(H_u)_{ij} = \frac{v^2}{2}d_i d^*_j\,. \label{case1333-HdHu}
\end{equation}
Both in the down and up-quark sectors, we have two generations of massless quarks 
and one generation of massive ones, with $m_b^2 = v^2 |\vec g|^2/2$ and $m_t^2 = v^2 |\vec d|^2/2$.

In general, the mass matrices are diagonalized through the rotations
\begin{equation}
	d_L^0 = V_{dL} d_L\,, \quad d_R^0 = V_{dR} d_R\,, \quad
	u_L^0 = V_{uL} u_L\,, \quad u_R^0 = V_{uR} u_R\,.\label{quark-rotations}
\end{equation}
which lead to the CKM matrix $\VCKM = V_{uL}^\dagger V_{dL}$ and 
the diagonal matrices $D_d = V_{dL}^\dagger M_d V_{dR}$ and $D_u = V_{uL}^\dagger M_u V_{uR}$. 
Let us assume that the non-zero masses correspond to the third eigenvector. 
Then the matrices $V_{dL}$ and $V_{uL}$ take the following generic form:
\begin{equation}
	V_{dL} = \frac{1}{\sqrt{|\vec g|^2}}\mtrx{\uparrow& \uparrow & \uparrow\\ \vec x_1&\vec x_2&\vec g\\ \downarrow& \downarrow & \downarrow}\,,\quad
	V_{uL} = \frac{1}{\sqrt{|\vec d|^2}}\mtrx{\uparrow& \uparrow & \uparrow\\ \vec y_1&\vec y_2&\vec d\\ \downarrow& \downarrow & \downarrow}\,.
	\label{case1333-VdLVuL}
\end{equation}
Here, for the down-quark sector, $\vec g = (g_1, g_2, g_3)$ and $\vec x_1$, $\vec x_2$ are the other two eigenvectors corresponding to zero eigenvalue, 
which cannot be uniquely defined. The same construction holds for the up-sector in terms of the vectors
$\vec d = (d_1, d_2, d_3)$ and $\vec y_1$, $\vec y_2$. 

Proceeding in a similar way with $G_d = M_d^\dagger M_d$ and $G_u = M_u^\dagger M_u$,
one finds $(G_d)_{ij} = |\vec g|^2 v_i^* v_j$ and $(G_u)_{ij} = |\vec d|^2 v_i v_j^*$.
One can then find the rotation matrices $V_{dR}$ and $V_{uR}$, inserting them back in the Yukawa lagrangian
and obtain the interaction patterns of the neutral scalar fields $\phi_a^0 = (h_a + i a_a)/\sqrt{2}$
with quark pairs:
\begin{equation}
	-{\cal L}_Y \supset \frac{\sqrt{2}m_b}{v^2} \bar{d}_{L} 
	\left( \begin{array}{ccc}0& 0 & 0\\ 0& 0 & 0 \\ 
		\vec{\phi^0}\cdot \vec u_1& \vec{\phi^0}\cdot \vec u_2 & \vec{\phi^0}\cdot \vec v\end{array}\right)  
	d_{R} + h.c.\,,\label{couplings-down}
\end{equation}
and a similar matrix for the up-quark sector.
Here, the two vectors $\vec u_1$ and $\vec u_2$, of norm $v$, are orthogonal to $\vec v$ and with each other.
Although the three vectors $\vec u_1$, $\vec u_2$, $\vec v$ are different for different vev alignments, 
they all share the key properties: the SM-like Higgs couples only to $\bar b b$, 
while all the additional neutral Higgses couple to $b\bar{d}$ and $b\bar{s}$.
In particular, for the vev alignment $B_1$ we recover the coupling patterns given in Eq.~\eqref{h-a-quark}.

{\bf Case 3-3-1-3}: $d_R$ are trivial singlets, $Q_L \sim \bm{3}$, $u_R \sim \bm{3}^*$.
The Yukawa matrices $\Gamma_a$ again come from the simple product of $\bm{3}^* \otimes \bm{3}$, 
which is then coupled to the three singlets $d_R$, each with its own coefficient:
\begin{equation}
	\Gamma_1 = \mtrx{g_1&g_2&g_3 \\ \cdot&\cdot&\cdot \\ \cdot&\cdot&\cdot}\,, \quad 
	\Gamma_2 = \mtrx{\cdot&\cdot&\cdot \\ g_1&g_2&g_3 \\ \cdot&\cdot&\cdot}\,, \quad 
	\Gamma_3 = \mtrx{\cdot&\cdot&\cdot \\ \cdot&\cdot&\cdot \\ g_1&g_2&g_3}\,.\label{case3313-Gamma}
\end{equation}
We arrive at $(H_d)_{ij} = |\vec g|^2v_i v^*_j/2$. 
again yielding two massless generations and a massive $b$-quark, with the same mass squared as before: $m_b^2 = v^2 |\vec g|^2/2$.
However, the quark rotation matrix is now governed by the vector of vevs rather than the coefficients:
\begin{equation}
	V_{dL} = \frac{1}{v}\mtrx{\uparrow& \uparrow & \uparrow\\ \vec u_1&\vec u_2&\vec v\\ \downarrow& \downarrow & \downarrow}\,, \label{Vdl-3313}
\end{equation}
while $V_{dR}$ takes the form similar to Eq.~\eqref{case1333-VdLVuL}.
The interactions of the neutral Higgses with quark pairs is now described by the matrix 
\begin{eqnarray}
	-{\cal L}_Y \supset \frac{\sqrt{2}m_b}{v^2} \bar{d}_{L} \left( \begin{array}{ccc}0& 0 & \vec{\phi^0}\cdot \vec u_1\\ 0& 0 & \vec{\phi^0}\cdot \vec u_2 \\ 0& 0 & \vec{\phi^0}\cdot \vec v\end{array}\right)  d_{R} + h.c.\,,\label{couplings-down-3313}
\end{eqnarray}
We again observe that the decays of $h$ and $a$ to $\bar b d$ and $\bar b s$ are unavoidable.

In the up-quark sector, we encounter the product $\bm{3}^* \otimes \bm{3}^* \otimes \bm{3}^*$,
which contains only one singlet. As a result, the entire up-quark sector now involves only one free parameters $d$.
Using the representation product rules listed in Appendix~\ref{appendix-group}, 
we obtain the following Yukawa matrices:
\begin{equation}
	\Delta_1 = d\mtrx{\cdot&\cdot&\cdot \\ \cdot&\cdot& -1 \\ \cdot& 1 &\cdot}\,, \quad 
	\Delta_2 = d\mtrx{\cdot&\cdot& 1 \\ \cdot&\cdot&\cdot \\  -1 &\cdot&\cdot}\,, \quad 
	\Delta_3 = d\mtrx{\cdot& -1 &\cdot \\  1 &\cdot&\cdot \\ \cdot&\cdot&\cdot}\,.\label{case3313-Delta}
\end{equation}
The mass matrix and its hermitean square are
\begin{equation}
	(M_u)_{ij} = -\frac{d}{\sqrt{2}}\epsilon_{ijk} v^*_k\,, \quad 
	(H_{u})_{ij} = \frac{|d|^2}{2}\left(v^2 \delta_{ij} - v^*_i v_j\right)\,.\label{up-Hd-3313}
\end{equation}
The mass spectrum is now different: one zero eigenvalue, with the eigenvector $(v_1^*, v_2^*, v_3^*)$
and two non-zero degenerate eigenvalues $m^2 = |d|^2 v^2/2$.
The Higgs-quark couplings now show a different pattern but the overall conclusion remains unchanged: 
$h$ and $a$ unavoidably decay to the quark pairs.

{\bf Case 3-3-3-1}, with $u_R$ being trivial singlets, while $Q_L \sim \bm{3}^*$, $d_R \sim \bm{3}$,
is similar to the previous one, with the structures in the up and down-quark sectors swapped.

Finally, {\bf case 3-3-1-3$^{(1)}$} and {\bf case 3-3-1-3$^{(3)}$}, 
with $Q_L \sim \bm{3}$ and $u_R \sim (\bm{3}^{(1)})^*$ or $u_R \sim (\bm{3}^{(3)})^*$,
involve new structures. 
The down sector remains the same as in case 3-3-1-3, while the up sector is now constructed 
from a different contraction of triplets:
\begin{equation}
	\Delta_1 = d\mtrx{\sqrt{2}\tau_\mp&\cdot&\cdot \\ \cdot&\cdot& \mp\tau_\pm \\ \cdot& \mp\tau_\pm &\cdot}\,, \quad 
	\Delta_2 = d\mtrx{\cdot&\cdot& \mp\tau_\pm \\ \cdot&\sqrt{2}\tau_\mp&\cdot \\  \mp\tau_\pm &\cdot&\cdot}\,, \quad 
	\Delta_3 = d\mtrx{\cdot& \mp\tau_\pm &\cdot \\  \mp\tau_\pm &\cdot&\cdot \\ \cdot&\cdot&\sqrt{2}\tau_\mp}\,,\label{case3313'-Delta}
\end{equation}
where $\tau_- = \sqrt{2(3-\sqrt{3})}$, $\tau_+ = \sqrt{2(3+\sqrt{3})}$, so that $\tau_-\tau_+ = \sqrt{24}$.
We now obtain a mass matrix with three non-zero masses, two of them being degenerate.
Although the analytic diagonalization for a generic vev alignment is cumbersome,
it is easily done for the alignments which are possible for the $\Sigma(36)$ potential.
The bottom line is the same: all new neutral Higgses couple to quark pairs, leading to decays of $h$ and $a$.


\begin{thebibliography}{99}

\bibitem{Cirelli:2024ssz}
M.~Cirelli, A.~Strumia and J.~Zupan,
[arXiv:2406.01705 [hep-ph]].

\bibitem{Deshpande:1977rw}
N.~G.~Deshpande and E.~Ma,
Phys. Rev. D \textbf{18}, 2574 (1978)
doi:10.1103/PhysRevD.18.2574

\bibitem{Ma:2006km}
E.~Ma,
Phys. Rev. D \textbf{73}, 077301 (2006)
doi:10.1103/PhysRevD.73.077301
[arXiv:hep-ph/0601225 [hep-ph]].

\bibitem{Barbieri:2006dq}
R.~Barbieri, L.~J.~Hall and V.~S.~Rychkov,
Phys. Rev. D \textbf{74}, 015007 (2006)
doi:10.1103/PhysRevD.74.015007
[arXiv:hep-ph/0603188 [hep-ph]].

\bibitem{LopezHonorez:2006gr}
L.~Lopez Honorez, E.~Nezri, J.~F.~Oliver and M.~H.~G.~Tytgat,
JCAP \textbf{02}, 028 (2007)
doi:10.1088/1475-7516/2007/02/028
[arXiv:hep-ph/0612275 [hep-ph]].

\bibitem{Arhrib:2013ela}
A.~Arhrib, Y.~L.~S.~Tsai, Q.~Yuan and T.~C.~Yuan,
JCAP \textbf{06}, 030 (2014)
doi:10.1088/1475-7516/2014/06/030
[arXiv:1310.0358 [hep-ph]].

\bibitem{Ilnicka:2015jba}
A.~Ilnicka, M.~Krawczyk and T.~Robens,
Phys. Rev. D \textbf{93}, no.5, 055026 (2016)
doi:10.1103/PhysRevD.93.055026
[arXiv:1508.01671 [hep-ph]].

\bibitem{Belyaev:2016lok}
A.~Belyaev, G.~Cacciapaglia, I.~P.~Ivanov, F.~Rojas-Abatte and M.~Thomas,
Phys. Rev. D \textbf{97}, no.3, 035011 (2018)
doi:10.1103/PhysRevD.97.035011
[arXiv:1612.00511 [hep-ph]].

\bibitem{Kalinowski:2018ylg}
J.~Kalinowski, W.~Kotlarski, T.~Robens, D.~Sokolowska and A.~F.~Zarnecki,
JHEP \textbf{12}, 081 (2018)
doi:10.1007/JHEP12(2018)081
[arXiv:1809.07712 [hep-ph]].

\bibitem{Belyaev:2018ext}
A.~Belyaev, T.~R.~Fernandez Perez Tomei, P.~G.~Mercadante, C.~S.~Moon, S.~Moretti, S.~F.~Novaes, L.~Panizzi, F.~Rojas and M.~Thomas,
Phys. Rev. D \textbf{99}, no.1, 015011 (2019)
doi:10.1103/PhysRevD.99.015011
[arXiv:1809.00933 [hep-ph]].

\bibitem{Kajiyama:2011gu}
Y.~Kajiyama, K.~Kannike and M.~Raidal,
Phys. Rev. D \textbf{85}, 033008 (2012)
doi:10.1103/PhysRevD.85.033008
[arXiv:1111.1270 [hep-ph]].

\bibitem{Ma:2007gq}
E.~Ma,
Phys. Lett. B \textbf{662}, 49-52 (2008)
doi:10.1016/j.physletb.2008.02.053
[arXiv:0708.3371 [hep-ph]].

\bibitem{Batell:2010bp}
B.~Batell,
Phys. Rev. D \textbf{83}, 035006 (2011)
doi:10.1103/PhysRevD.83.035006
[arXiv:1007.0045 [hep-ph]].

\bibitem{Ivanov:2012hc}
I.~P.~Ivanov and V.~Keus,
Phys. Rev. D \textbf{86}, 016004 (2012)
doi:10.1103/PhysRevD.86.016004
[arXiv:1203.3426 [hep-ph]].

\bibitem{Belanger:2012vp}
G.~Belanger, K.~Kannike, A.~Pukhov and M.~Raidal,
JCAP \textbf{04}, 010 (2012)
doi:10.1088/1475-7516/2012/04/010
[arXiv:1202.2962 [hep-ph]].

\bibitem{Belanger:2014bga}
G.~B\'elanger, K.~Kannike, A.~Pukhov and M.~Raidal,
JCAP \textbf{06}, 021 (2014)
doi:10.1088/1475-7516/2014/06/021
[arXiv:1403.4960 [hep-ph]].

\bibitem{Yaguna:2019cvp}
C.~E.~Yaguna and \'O.~Zapata,
JHEP \textbf{03}, 109 (2020)
doi:10.1007/JHEP03(2020)109
[arXiv:1911.05515 [hep-ph]].

\bibitem{Belanger:2020hyh}
G.~B\'elanger, A.~Pukhov, C.~E.~Yaguna and \'O.~Zapata,
JHEP \textbf{09}, 030 (2020)
doi:10.1007/JHEP09(2020)030
[arXiv:2006.14922 [hep-ph]].

\bibitem{Yaguna:2021vhb}
C.~E.~Yaguna and \'O.~Zapata,
JHEP \textbf{10}, 185 (2021)
doi:10.1007/JHEP10(2021)185
[arXiv:2106.11889 [hep-ph]].

\bibitem{Yaguna:2021rds}
C.~E.~Yaguna and \'O.~Zapata,
Phys. Rev. D \textbf{105}, no.9, 095026 (2022)
doi:10.1103/PhysRevD.105.095026
[arXiv:2112.07020 [hep-ph]].

\bibitem{Belanger:2021lwd}
G.~Belanger, A.~Mjallal and A.~Pukhov,
Phys. Rev. D \textbf{105}, no.3, 035018 (2022)
doi:10.1103/PhysRevD.105.035018
[arXiv:2108.08061 [hep-ph]].

\bibitem{Belanger:2022esk}
G.~B\'elanger, A.~Pukhov, C.~E.~Yaguna and \'O.~Zapata,
JHEP \textbf{03}, 100 (2023)
doi:10.1007/JHEP03(2023)100
[arXiv:2212.07488 [hep-ph]].

\bibitem{Ivanov:2015mwl}
I.~P.~Ivanov and J.~P.~Silva,
Phys. Rev. D \textbf{93}, no.9, 095014 (2016)
doi:10.1103/PhysRevD.93.095014
[arXiv:1512.09276 [hep-ph]].

\bibitem{Ivanov:2018srm}
I.~P.~Ivanov and M.~Laletin,
JCAP \textbf{05}, 032 (2019)
doi:10.1088/1475-7516/2019/05/032
[arXiv:1812.05525 [hep-ph]].

\bibitem{Belanger:2011ww}
G.~Belanger and J.~C.~Park,
JCAP \textbf{03}, 038 (2012)
doi:10.1088/1475-7516/2012/03/038
[arXiv:1112.4491 [hep-ph]].

\bibitem{Aoki:2012ub}
M.~Aoki, M.~Duerr, J.~Kubo and H.~Takano,
Phys. Rev. D \textbf{86}, 076015 (2012)
doi:10.1103/PhysRevD.86.076015
[arXiv:1207.3318 [hep-ph]].

\bibitem{Bhattacharya:2016ysw}
S.~Bhattacharya, P.~Poulose and P.~Ghosh,
JCAP \textbf{04}, 043 (2017)
doi:10.1088/1475-7516/2017/04/043
[arXiv:1607.08461 [hep-ph]].

\bibitem{Hernandez-Sanchez:2020aop}
J.~Hernandez-Sanchez, V.~Keus, S.~Moretti, D.~Rojas-Ciofalo and D.~Sokolowska,
[arXiv:2012.11621 [hep-ph]].

\bibitem{Boto:2024tzp}
R.~Boto, P.~N.~Figueiredo, J.~C.~Rom\~ao and J.~P.~Silva,
JHEP \textbf{11}, 108 (2024)
doi:10.1007/JHEP11(2024)108
[arXiv:2407.15933 [hep-ph]].

\bibitem{Adulpravitchai:2011ei}
A.~Adulpravitchai, B.~Batell and J.~Pradler,
Phys. Lett. B \textbf{700}, 207-216 (2011)
doi:10.1016/j.physletb.2011.04.015
[arXiv:1103.3053 [hep-ph]].

\bibitem{Lovrekovic:2012bz}
I.~Lovrekovic,
[arXiv:1212.1145 [hep-ph]].

\bibitem{Ivanov:2012ry}
I.~P.~Ivanov and E.~Vdovin,
Phys. Rev. D \textbf{86}, 095030 (2012)
doi:10.1103/PhysRevD.86.095030
[arXiv:1206.7108 [hep-ph]].

\bibitem{Ivanov:2012fp}
I.~P.~Ivanov and E.~Vdovin,
Eur. Phys. J. C \textbf{73}, no.2, 2309 (2013)
doi:10.1140/epjc/s10052-013-2309-x
[arXiv:1210.6553 [hep-ph]].

\bibitem{Ivanov:2014doa}
I.~P.~Ivanov and C.~C.~Nishi,
JHEP \textbf{01}, 021 (2015)
doi:10.1007/JHEP01(2015)021
[arXiv:1410.6139 [hep-ph]].

\bibitem{Yang:2024bys}
Y.~Yang and I.~P.~Ivanov,
Phys. Rev. D \textbf{110}, no.1, 015001 (2024)
doi:10.1103/PhysRevD.110.015001
[arXiv:2401.03264 [hep-ph]].

\bibitem{Alguero:2023zol}
G.~Alguero, G.~Belanger, F.~Boudjema, S.~Chakraborti, A.~Goudelis, S.~Kraml, A.~Mjallal and A.~Pukhov,
Comput. Phys. Commun. \textbf{299}, 109133 (2024)
doi:10.1016/j.cpc.2024.109133
[arXiv:2312.14894 [hep-ph]].

\bibitem{Planck:2018vyg}
N.~Aghanim \textit{et al.} [Planck],
Astron. Astrophys. \textbf{641}, A6 (2020)
[erratum: Astron. Astrophys. \textbf{652}, C4 (2021)]
doi:10.1051/0004-6361/201833910
[arXiv:1807.06209 [astro-ph.CO]].

\bibitem{XENON:2018voc}
E.~Aprile \textit{et al.} [XENON],
Phys. Rev. Lett. \textbf{121}, no.11, 111302 (2018)
doi:10.1103/PhysRevLett.121.111302
[arXiv:1805.12562 [astro-ph.CO]].

\bibitem{LZ:2022lsv}
J.~Aalbers \textit{et al.} [LZ],
Phys. Rev. Lett. \textbf{131}, no.4, 041002 (2023)
doi:10.1103/PhysRevLett.131.041002
[arXiv:2207.03764 [hep-ex]].

\bibitem{LZCollaboration:2024lux}
J.~Aalbers \textit{et al.} [LZ],
[arXiv:2410.17036 [hep-ex]].

\bibitem{Ivanov:2017dad}
I.~P.~Ivanov,
Prog. Part. Nucl. Phys. \textbf{95}, 160-208 (2017)
doi:10.1016/j.ppnp.2017.03.001
[arXiv:1702.03776 [hep-ph]].

\bibitem{Segre:1978ji}
G.~Segre and H.~A.~Weldon,
Phys. Lett. B \textbf{83}, 351-354 (1979)
doi:10.1016/0370-2693(79)91125-0

\bibitem{Degee:2012sk}
A.~Degee, I.~P.~Ivanov and V.~Keus,
JHEP \textbf{02}, 125 (2013)
doi:10.1007/JHEP02(2013)125
[arXiv:1211.4989 [hep-ph]].

\bibitem{Branco:1983tn}
G.~C.~Branco, J.~M.~Gerard and W.~Grimus,
Phys. Lett. B \textbf{136}, 383-386 (1984)
doi:10.1016/0370-2693(84)92024-0

\bibitem{deMedeirosVarzielas:2011zw}
I.~de Medeiros Varzielas and D.~Emmanuel-Costa,
Phys. Rev. D \textbf{84}, 117901 (2011)
doi:10.1103/PhysRevD.84.117901
[arXiv:1106.5477 [hep-ph]].

\bibitem{deMedeirosVarzielas:2012rxf}
I.~de Medeiros Varzielas,
JHEP \textbf{08}, 055 (2012)
doi:10.1007/JHEP08(2012)055
[arXiv:1205.3780 [hep-ph]].

\bibitem{Ivanov:2013nla}
I.~P.~Ivanov and L.~Lavoura,
Eur. Phys. J. C \textbf{73}, no.4, 2416 (2013)
doi:10.1140/epjc/s10052-013-2416-8
[arXiv:1302.3656 [hep-ph]].

\bibitem{Leurer:1992wg}
M.~Leurer, Y.~Nir and N.~Seiberg,
Nucl. Phys. B \textbf{398}, 319-342 (1993)
doi:10.1016/0550-3213(93)90112-3
[arXiv:hep-ph/9212278 [hep-ph]].

\bibitem{GonzalezFelipe:2014mcf}
R.~Gonz\'alez Felipe, I.~P.~Ivanov, C.~C.~Nishi, H.~Ser\^odio and J.~P.~Silva,
Eur. Phys. J. C \textbf{74}, no.7, 2953 (2014)
doi:10.1140/epjc/s10052-014-2953-9
[arXiv:1401.5807 [hep-ph]].

\bibitem{GonzalezFelipe:2013xok}
R.~Gonz\'alez Felipe, H.~Ser\^odio and J.~P.~Silva,
Phys. Rev. D \textbf{87}, no.5, 055010 (2013)
doi:10.1103/PhysRevD.87.055010
[arXiv:1302.0861 [hep-ph]].

\bibitem{Bree:2023ojl}
I.~Bree, S.~Carrolo, J.~C.~Romao and J.~P.~Silva,
Eur. Phys. J. C \textbf{83}, no.4, 292 (2023)
doi:10.1140/epjc/s10052-023-11463-5
[arXiv:2301.04676 [hep-ph]].

\bibitem{Varzielas:2012nn}
I.~de Medeiros Varzielas, D.~Emmanuel-Costa and P.~Leser,
Phys. Lett. B \textbf{716}, 193-196 (2012)
doi:10.1016/j.physletb.2012.08.008
[arXiv:1204.3633 [hep-ph]].

\bibitem{Varzielas:2013eta}
I.~de Medeiros Varzielas and D.~Pidt,
JHEP \textbf{11}, 206 (2013)
doi:10.1007/JHEP11(2013)206
[arXiv:1307.6545 [hep-ph]].

\bibitem{Kalinowski:2021lvw}
J.~Kalinowski, W.~Kotlarski, M.~N.~Rebelo and I.~de Medeiros Varzielas,
JHEP \textbf{02}, 231 (2023)
doi:10.1007/JHEP02(2023)231
[arXiv:2112.12699 [hep-ph]].

\bibitem{Ivanov:2011ae}
I.~P.~Ivanov, V.~Keus and E.~Vdovin,
J. Phys. A \textbf{45}, 215201 (2012)
doi:10.1088/1751-8113/45/21/215201
[arXiv:1112.1660 [math-ph]].

\bibitem{Grimus:2010ak}
W.~Grimus and P.~O.~Ludl,
J. Phys. A \textbf{43}, 445209 (2010)
doi:10.1088/1751-8113/43/44/445209
[arXiv:1006.0098 [hep-ph]].

\bibitem{Hagedorn:2013nra}
C.~Hagedorn, A.~Meroni and L.~Vitale,
J. Phys. A \textbf{47}, 055201 (2014)
doi:10.1088/1751-8113/47/5/055201
[arXiv:1307.5308 [hep-ph]].

\bibitem{Maniatis:2006fs}
M.~Maniatis, A.~von Manteuffel, O.~Nachtmann and F.~Nagel,
Eur. Phys. J. C \textbf{48}, 805-823 (2006)
doi:10.1140/epjc/s10052-006-0016-6
[arXiv:hep-ph/0605184 [hep-ph]].

\bibitem{Aoki:2023lbz}
M.~Aoki, L.~Biermann, C.~Borschensky, I.~P.~Ivanov, M.~M\"uhlleitner and H.~Shibuya,
JHEP \textbf{02}, 232 (2024)
doi:10.1007/JHEP02(2024)232
[arXiv:2308.04141 [hep-ph]].

\bibitem{deMedeirosVarzielas:2015amz}
I.~de Medeiros Varzielas,
JHEP \textbf{08}, 157 (2015)
doi:10.1007/JHEP08(2015)157
[arXiv:1507.00338 [hep-ph]].

\bibitem{deMedeirosVarzielas:2021zqs}
I.~de Medeiros Varzielas, I.~P.~Ivanov and M.~Levy,
Eur. Phys. J. C \textbf{81}, no.10, 918 (2021)
doi:10.1140/epjc/s10052-021-09681-w
[arXiv:2107.08227 [hep-ph]].

\bibitem{Khater:2021wcx}
W.~Khater, A.~Kun\v{c}inas, O.~M.~Ogreid, P.~Osland and M.~N.~Rebelo,
JHEP \textbf{01}, 120 (2022)
doi:10.1007/JHEP01(2022)120
[arXiv:2108.07026 [hep-ph]].

\bibitem{Kuncinas:2022whn}
A.~Kun\v{c}inas, O.~M.~Ogreid, P.~Osland and M.~N.~Rebelo,
Phys. Rev. D \textbf{106}, no.7, 075002 (2022)
doi:10.1103/PhysRevD.106.075002
[arXiv:2204.05684 [hep-ph]].

\bibitem{ATLAS:2023tkt}
G.~Aad \textit{et al.} [ATLAS],
Phys. Lett. B \textbf{842}, 137963 (2023)
doi:10.1016/j.physletb.2023.137963
[arXiv:2301.10731 [hep-ex]].

\bibitem{Djouadi:2005gj}
A.~Djouadi,
Phys. Rept. \textbf{459}, 1-241 (2008)
doi:10.1016/j.physrep.2007.10.005
[arXiv:hep-ph/0503173 [hep-ph]].

\bibitem{Cao:2007rm}
Q.~H.~Cao, E.~Ma and G.~Rajasekaran,
Phys. Rev. D \textbf{76}, 095011 (2007)
doi:10.1103/PhysRevD.76.095011
[arXiv:0708.2939 [hep-ph]].

\bibitem{Arhrib:2012ia}
A.~Arhrib, R.~Benbrik and N.~Gaur,
Phys. Rev. D \textbf{85}, 095021 (2012)
doi:10.1103/PhysRevD.85.095021
[arXiv:1201.2644 [hep-ph]].

\bibitem{Swiezewska:2012eh}
B.~Swiezewska and M.~Krawczyk,
Phys. Rev. D \textbf{88}, no.3, 035019 (2013)
doi:10.1103/PhysRevD.88.035019
[arXiv:1212.4100 [hep-ph]].

\bibitem{Krawczyk:2013jta}
M.~Krawczyk, D.~Sokolowska, P.~Swaczyna and B.~Swiezewska,
JHEP \textbf{09}, 055 (2013)
doi:10.1007/JHEP09(2013)055
[arXiv:1305.6266 [hep-ph]].

\bibitem{ParticleDataGroup:2024cfk}
S.~Navas \textit{et al.} [Particle Data Group],
Phys. Rev. D \textbf{110}, no.3, 030001 (2024)
doi:10.1103/PhysRevD.110.030001

\end{thebibliography}
\end{document}